\documentclass[twocolumn,amsmath,amssymb,prd]{revtex4}

\def\al{\alpha}
\def\be{\beta}
\def\ga{\gamma}
\def\de{\delta}
\def\ep{\epsilon}

\def\ze{\zeta}
\def\et{\eta}
\def\th{\theta}

\def\la{\lambda}

\def\rh{\rho}

\def\si{\sigma}
\def\vs{\varsigma}

\def\ch{\chi}
\def\ps{\psi}
\def\om{\omega}
\def\Ga{\Gamma}
\def\De{\Delta}

\def\Si{\Sigma}

\def\cH{{\cal H}}
\def\cl{{\cal L}}

\def\fr#1#2{{{#1}\over{#2}}}
\def\frac#1#2{{\textstyle{{#1}\over{#2}}}}
\def\half{{\textstyle{1\over 2}}}
\def\ol{\overline}
\def\prt{\partial}

\def\lsim{\mathrel{\rlap{\lower4pt\hbox{\hskip1pt$\sim$}}
    \raise1pt\hbox{$<$}}}
\def\gsim{\mathrel{\rlap{\lower4pt\hbox{\hskip1pt$\sim$}}
    \raise1pt\hbox{$>$}}}

\def\etal{{\it et al.}}

\def\vev#1{\langle {#1}\rangle}

\newcommand{\beq}{\begin{equation}}
\newcommand{\eeq}{\end{equation}}
\newcommand{\bea}{\begin{eqnarray}}
\newcommand{\eea}{\end{eqnarray}}
\newcommand{\rf}[1]{(\ref{#1})}

\def\nn{\nonumber}

\def\psb{\ol\ps{}}

\def\mbf#1{\boldsymbol #1}

\def\Q{\mathcal Q}

\def\pvec{\mbf p}
\def\gavec{\mbf\ga}

\def\Qhat{\widehat\Q}

\def\X{X}
\def\Y{Y}
\def\Z{Z}
\def\Xhat{\widehat\X}
\def\Yhat{\widehat\Y}
\def\Zhat{\widehat\Z}

\def\codt{\cos{\om_\oplus T_\oplus}}
\def\sodt{\sin{\om_\oplus T_\oplus}}
\def\ctodt{\cos{2\om_\oplus T_\oplus}}
\def\stodt{\sin{2\om_\oplus T_\oplus}}

\def\cmtemplate#1#2#3#4{{#1}^{#3}_{#4}}
\def\mfcm#1#2{\cmtemplate{m}{#1}{#2}{5}}
\def\acm#1#2{\cmtemplate{a}{#1}{#2}{}}
\def\bcm#1#2{\cmtemplate{b}{#1}{#2}{}}
\def\ccm#1#2{\cmtemplate{c}{#1}{#2}{}}
\def\dcm#1#2{\cmtemplate{d}{#1}{#2}{}}
\def\ecm#1#2{\cmtemplate{e}{#1}{#2}{}}
\def\fcm#1#2{\cmtemplate{f}{#1}{#2}{}}
\def\gcm#1#2{\cmtemplate{g}{#1}{#2}{}}
\def\Hcm#1#2{\cmtemplate{H}{#1}{#2}{}}

\def\ctemplate#1#2#3#4{{#1}^{(#2)#3}_{#4}}
\def\mc#1#2{\ctemplate{m}{#1}{#2}{}}
\def\mfc#1#2{\ctemplate{m}{#1}{#2}{5}}
\def\ac#1#2{\ctemplate{a}{#1}{#2}{}}
\def\bc#1#2{\ctemplate{b}{#1}{#2}{}}
\def\cc#1#2{\ctemplate{c}{#1}{#2}{}}
\def\dc#1#2{\ctemplate{d}{#1}{#2}{}}
\def\ec#1#2{\ctemplate{e}{#1}{#2}{}}
\def\fc#1#2{\ctemplate{f}{#1}{#2}{}}
\def\gc#1#2{\ctemplate{g}{#1}{#2}{}}
\def\Hc#1#2{\ctemplate{H}{#1}{#2}{}}

\def\mcf#1#2{\ctemplate{m}{#1}{#2}{F}}
\def\mfcf#1#2{\ctemplate{m}{#1}{#2}{5F}}
\def\acf#1#2{\ctemplate{a}{#1}{#2}{F}}
\def\bcf#1#2{\ctemplate{b}{#1}{#2}{F}}
\def\ccf#1#2{\ctemplate{c}{#1}{#2}{F}}
\def\dcf#1#2{\ctemplate{d}{#1}{#2}{F}}
\def\ecf#1#2{\ctemplate{e}{#1}{#2}{F}}
\def\fcf#1#2{\ctemplate{f}{#1}{#2}{F}}
\def\gcf#1#2{\ctemplate{g}{#1}{#2}{F}}
\def\Hcf#1#2{\ctemplate{H}{#1}{#2}{F}}

\def\mcpf#1#2{\ctemplate{m}{#1}{#2}{\prt F}}
\def\mfcpf#1#2{\ctemplate{m}{#1}{#2}{5\prt F}}
\def\acpf#1#2{\ctemplate{a}{#1}{#2}{\prt F}}
\def\bcpf#1#2{\ctemplate{b}{#1}{#2}{\prt F}}

\def\Hcpf#1#2{\ctemplate{H}{#1}{#2}{\prt F}}

\def\cmtemplate#1#2#3#4{{#1}^{#3}_{#4}}

\def\acmw#1#2#3{\cmtemplate{a}{#1}{#2}{{#3}}}
\def\bcmw#1#2#3{\cmtemplate{b}{#1}{#2}{{#3}}}
\def\ccmw#1#2#3{\cmtemplate{c}{#1}{#2}{{#3}}}
\def\dcmw#1#2#3{\cmtemplate{d}{#1}{#2}{{#3}}}
\def\ecmw#1#2#3{\cmtemplate{e}{#1}{#2}{{#3}}}

\def\gcmw#1#2#3{\cmtemplate{g}{#1}{#2}{{#3}}}
\def\Hcmw#1#2#3{\cmtemplate{H}{#1}{#2}{{#3}}}

\def\ctemplate#1#2#3#4{{#1}^{(#2)#3}_{#4}}
\def\mcw#1#2#3{\ctemplate{m}{#1}{#2}{{#3}}}

\def\acw#1#2#3{\ctemplate{a}{#1}{#2}{{#3}}}
\def\bcw#1#2#3{\ctemplate{b}{#1}{#2}{{#3}}}
\def\ccw#1#2#3{\ctemplate{c}{#1}{#2}{{#3}}}
\def\dcw#1#2#3{\ctemplate{d}{#1}{#2}{{#3}}}
\def\ecw#1#2#3{\ctemplate{e}{#1}{#2}{{#3}}}

\def\gcw#1#2#3{\ctemplate{g}{#1}{#2}{{#3}}}
\def\Hcw#1#2#3{\ctemplate{H}{#1}{#2}{{#3}}}

\def\mcfw#1#2#3{\ctemplate{m}{#1}{#2}{F{,#3}}}

\def\acfw#1#2#3{\ctemplate{a}{#1}{#2}{F{,#3}}}
\def\bcfw#1#2#3{\ctemplate{b}{#1}{#2}{F{,#3}}}
\def\ccfw#1#2#3{\ctemplate{c}{#1}{#2}{F{,#3}}}
\def\dcfw#1#2#3{\ctemplate{d}{#1}{#2}{F{,#3}}}
\def\ecfw#1#2#3{\ctemplate{e}{#1}{#2}{F{,#3}}}

\def\gcfw#1#2#3{\ctemplate{g}{#1}{#2}{F{,#3}}}
\def\Hcfw#1#2#3{\ctemplate{H}{#1}{#2}{F{,#3}}}

\def\mn{{\mu\nu}}
\def\ma{{\mu\al}}
\def\mna{{\mu\nu\al}}
\def\ab{{\al\be}}
\def\bec{{\be\ga}}
\def\mab{{\mu\al\be}}
\def\mnab{{\mu\nu\al\be}}
\def\abc{{\al\be\ga}}

\def\mabc{{\mu\al\be\ga}}
\def\mnabc{{\mu\nu\al\be\ga}}
\def\abcd{{\al\be\ga\de}}
\def\va{{\vs\al}}
\def\vab{{\vs\al\be}}
\def\vabc{{\vs\al\be\ga}}
\def\vabcd{{\vs\al\be\ga\de}}

\def\m{m_\ps}
\def\mw{m_w}
\def\Z{\si}

\def\quar{\frac 1 4}

\def\ens{E_{n,s}}
\def\enms{E_{n,-s}}

\def\atw#1#2{{\widetilde a}_{#1}^{#2}}
\def\btw#1#2{{\widetilde b}_{#1}^{#2}}
\def\bftw#1#2{{\widetilde b}_{F,#1}^{#2}}
\def\mftw#1#2{{\widetilde m}_{F,#1}^{#2}}
\def\atws#1#2{{\widetilde a}_{#1}^{*#2}}
\def\btws#1#2{{\widetilde b}_{#1}^{*#2}}
\def\bftws#1#2{{\widetilde b}_{F,#1}^{*#2}}
\def\mftws#1#2{{\widetilde m}_{F,#1}^{*#2}}

\begin{document}
\title{
Lorentz-violating spinor electrodynamics and Penning traps 
}

\author{Yunhua Ding and V.\ Alan Kosteleck\'y}

\affiliation{Physics Department, Indiana University, 
Bloomington, Indiana 47405, USA}

\date{IUHET 601, August 2016}

\begin{abstract}
The prospects are explored for testing Lorentz- and CPT-violating
quantum electrodynamics in experiments with Penning traps.
We present the Lagrange density 
of Lorentz-violating spinor electrodynamics
with operators of mass dimensions up to six,
and we discuss some of its properties.
The theory is used to derive Lorentz- and CPT-violating perturbative shifts
of the energy levels of a particle confined to a Penning trap. 
Observable signals are discussed 
for trapped electrons, positrons, protons, and antiprotons.
Existing experimental measurements on anomaly frequencies
are used to extract new or improved bounds
on numerous coefficients for Lorentz and CPT violation,
using sidereal variations of observables
and comparisons between particles and antiparticles.
\end{abstract}

\maketitle

\section{Introduction}
\label{Introduction}

A powerful approach to investigating the fundamental properties
of a stable particle is to trap it for an extended period,
which allows probing it in detail. 
Electromagnetic traps operate
by taking advantage of a charge or magnetic moment
to confine the particle using a suitable field configuration.
For charged particles,
the Penning trap is a standard tool.
An idealized Penning trap involves a uniform magnetic field
bounding the particle motion in the perpendicular plane,
together with a quadrupole electrostatic field
preventing escape along the axis.
Penning traps can be used to achieve impressive sensitivities
to properties of fundamental particles,
as originally demonstrated by Dehmelt {\it et al.}
in measurements of the electron $g-2$ factor
and in a comparison of the electron and positron $g$ factors 
to parts in a trillion
\cite{de86,vd87}.

The high sensitivity offered by experiments with Penning traps 
implies they are well suited to precision studies of fundamental symmetries.
This includes the foundational Lorentz and CPT invariances of relativity. 
Studies of these invariances
have undergone a renaissance in recent years,
following the observation that
tiny violations of Lorentz symmetry could emerge 
in models unifying gravity with quantum physics
such as string theory
\cite{ksp}.
The potential opportunity to detect experimentally 
a physical effect arising from the Planck scale 
$M_P \simeq 10^{19}$~GeV
has stimulated many new high-precision searches for relativity violations
across various subfields of physics
\cite{tables}.
Here,
we advance this active area of research 
by investigating the prospects for searches for Lorentz and CPT violation
via spectroscopy of particles in Penning traps.

One possible approach to studying Lorentz and CPT violation
is to propose a specific model and investigate its implications.
However,
given the current absence of compelling experimental evidence 
for Lorentz and CPT violation,
it is advantageous to work within a general and realistic framework
allowing for all possible types of violations,
thereby offering a comprehensive treatment for prospective searches.

A general methodology for studying tiny signals 
arising as suppressed effects from an inaccessible sector 
is provided by effective field theory
\cite{sw}. 
For Lorentz violation,
the comprehensive realistic effective field theory 
can be constructed from General Relativity 
and the Standard Model of particle physics
by adding to the action all Lorentz-violating operators,
each contracted with a controlling coefficient
that maintains coordinate independence of the physics 
\cite{ck,akgrav}.
In this framework,
known as the Standard-Model Extension (SME),
Lorentz-violating operators of larger mass dimension $d$
can be interpreted as higher-order effects 
appearing in the low-energy limit.
The SME also describes general CPT-violating physics 
because the breaking of CPT symmetry
in the context of effective field theory
is accompanied by Lorentz violation 
\cite{ck,owg}.
Restricting attention to operators of renormalizable dimension $d\leq 4$
yields the minimal SME,
which in Minkowski spacetime is power-counting renormalizable.
The experimental implications of any desired specific model
that is compatible with effective field theory 
can be obtained from the SME framework
by matching the model parameters 
to a suitable subset of the SME coefficients
and adopting the corresponding experimental constraints
\cite{tables,reviews}.

The minimal SME reveals that Lorentz and CPT violation
can induce a variety of subtle but measurable effects in experiments 
studying the anomalous magnetic moment or charge-to-mass ratio
of a particle confined to a Penning trap
\cite{bkr97,bkr98}.
These effects include shifts in the anomaly and cyclotron frequencies
that can differ between particles and antiparticles
and that can vary with sidereal time.
Experimental searches for these SME effects that have been published to date
compare the electron and positron anomaly frequencies
\cite{de99},
constrain sidereal signals in the electron anomaly
and cyclotron frequencies
\cite{mi99},
and measure the cyclotron frequency 
of the H$^-$ ion relative to the antiproton
\cite{ga99,ul15}.
On the theory side,
several treatments have been given 
of Penning-trap sensitivities to Lorentz and CPT signals
both in the minimal SME and also
for certain nonminimal SME terms involving interactions at $d=5$
\cite{bkr97,bkr98,fr12,ar15,ar16}.

In the present work,
we further the theoretical basis for studies of Lorentz and CPT symmetry
in Penning traps
by developing the relevant nonminimal sector of the SME,
studying its properties,
and determining its predicted signals
for trapped electrons, positrons, protons, and antiprotons.
The recent characterization and enumeration of effects 
arising when a Dirac fermion propagates 
in the presence of Lorentz-violating operators 
of arbitrary mass dimension $d$
\cite{km13}
provides a partial guide for investigations of nonminimal effects
on particles in a Penning trap.
However,
the interactions of the particle 
with the electromagnetic fields in the trap  
can introduce additional types of nonminimal Lorentz violations
beyond those associated with propagation,
and these additional effects lack a systematic treatment 
in the literature to date.
One goal of this work is to address this gap,
by presenting and investigating
the explicit Lagrange density 
for Lorentz-violating spinor electrodynamics
that describes the behavior of a fermion coupled to the electromagnetic field
in the presence of both minimal and nonminimal Lorentz and CPT violation
with $d\leq 6$.
More generally,
investigations of nonminimal SME effects 
are of significance to various aspects of Lorentz and CPT violation,
ranging from phenomenological implications of specific models
involving noncommutative quantum field theory
\cite{ha00,chklo01}
or supersymmetry 
\cite{susy}
to more formal issues
such as the stability and causality 
of Lorentz-violating quantum field theories
\cite{causality}
or their mathematical foundations in Riemann-Finsler geometry 
\cite{finsler}.
The theoretical aspects discussed here are thus
of relevance beyond the immediate implications for experiments. 

Another major goal of this work is to establish specific observables 
for both minimal and nonminimal Lorentz and CPT violation
that are relevant to existing or near-future experiments
on particles in Penning traps.
We use perturbation theory to determine
the dominant Lorentz- and CPT-violating shifts 
in the anomaly and cyclotron frequencies
of electrons, positrons, protons, and antiprotons.
Armed with this information,
we revisit published experimental studies of Lorentz and CPT symmetry
with Penning traps
\cite{de99,mi99}
and extract some additional constraints.
The perturbative analysis also reveals bounds on SME coefficients
arising from other data,
including measurements of the electron anomaly frequency
\cite{ha11}
and of the proton and antiproton magnetic moments
\cite{di12,di13,mo14},
and it permits identification of potential signals
in forthcoming experiments
with positrons
\cite{fo15}
and antiprotons
\cite{base,chip}.
Here,
we extract constraints on SME coefficients from available data 
and provide tools for the analysis of future experiments. 
The results are complementary to existing and proposed studies
of Lorentz and CPT violation
involving measurements of the muon anomalous magnetic moment
via magnetic confinement in a ring accelerator 
\cite{muon,gkv14},
and more generally to constraints on nonminimal coefficients
in the electron and proton sectors from experiments
on hydrogen, antihydrogen, and related systems
\cite{kv15}. 

This work is organized as follows.
In Sec.\ \ref{Theory},
we present and investigate some properties of
quantum electrodynamics (QED) with Lorentz- and CPT-violating 
operators of dimensions $d\leq 6$.
The Lagrange density is given in 
Sec.\ \ref{Lagrange density},
along with its relation to some special models in the literature.
The issue of field redefinitions and physical observables is tackled in
Sec.\ \ref{Field redefinitions}.
We consider gauge-covariant invertible fermion redefinitions in
Sec.\ \ref{Fermion redefinitions},
tabulating the effects of each possibility.
In Sec.\ \ref{Absorption},
the issue of absorbing a given fermion coupling
to the electromagnetic field 
into other terms in the Lagrange density is addressed.
The special case of field redefinitions and observables
in the presence of a constant electromagnetic field,
which is of prime importance in the context of Penning traps,
is treated in 
Sec.\ \ref{Scenario}.
The experimental observables are affected 
by the noninertial nature of any laboratory frame
on the Earth,
and the necessary generic frame changes to convert results
to the canonical Sun-centered frame are described in  
Sec.\ \ref{Frame changes}.

We next turn in
Sec.\ \ref{Application to Penning traps}
to applications of the theory
to Penning-trap experiments.
Theoretical aspects of this subject are addressed in
Sec.\ \ref{Theory for Penning trap}
in the context of trapped electrons, positrons,
protons, and antiprotons.
We begin in Sec.\ \ref{Perturbative energy shift}
by deriving the dominant Lorentz- and CPT-violating perturbative shifts
of the energy levels of the trapped fermion,
and then turn in
Sec.\ \ref{Cyclotron and anomaly frequencies}
to a derivation of the effects 
on the cyclotron and anomaly frequencies
of trapped particles.
The experimental implications of these results are the subject of
Sec.\ \ref{Experiments}.
Some conceptual issues for experimental analyses
are considered in Sec.\ \ref{Concepts}.
In Sec.\ \ref{Sensitivities},
we investigate existing and prospective signals for experiments,
and we use the results together with published data
to extract bounds on various SME coefficients,
including some that were previously unconstrained.
In Sec.\ \ref{Summary},
we summarize the work and provide some outlook.
Finally,
Appendix \ref{Energy shift} contains some detailed results
for the perturbative Lorentz- and CPT-violating energy shifts.
The notation and conventions in this work
follow those of Ref.\ \cite{km13},
except as otherwise indicated.
In particular, 
we work in natural units with $c = \hbar = 1$.

\section{Theory}
\label{Theory}

In this section,
we present the Lagrange density for the fermion sector 
of Lorentz-violating QED,
incorporating operators with $d\leq 6$.
The procedure for using field redefinitions
to identify physical observables is discussed,
and the effects of a key set of redefinitions are tabulated.
Particular attention is paid to the special case of constant external field
relevant to many experimental configurations,
including those using a Penning trap discussed in this work.

\subsection{Lagrange density}
\label{Lagrange density}

The Lorentz-violating QED for a single Dirac fermion field $\ps$
of mass $\m$ and charge $q$ coupled to the photon field $A_\mu$
can be constructed by adding to the action of conventional QED
all terms that preserve U(1) gauge invariance 
formed from contractions of Lorentz-violating operators
with coefficients for Lorentz violation
\cite{ck}.
The coefficients can be viewed as background fields
that induce coordinate-independent Lorentz- and CPT-violating effects.
For operators of arbitrary mass dimension $d$,
the fermion sector of this theory
can be specified via a Lagrange density of the form
\beq
\cl_\ps =
\half \psb (\ga^\mu i D_\mu - \m + \Qhat) \ps + {\rm h.c.} , 
\label{fermlag}
\eeq
where $\Qhat$ is a $4\times4$ 
spinor matrix depending on the coefficients for Lorentz violation,
the covariant derivative
$iD_\al$
and the electromagnetic field strength 
$F_\ab \equiv \prt_\al A_\be - \prt_\be A_\al$.
The covariant derivative acting on the spinor takes the standard form 
$iD_\al \ps = (i \prt_\al - q A_\al) \ps$.
Note that $\Qhat$ satisfies the hermiticity condition
$\Qhat = \ga_0 \Qhat^\dag\ga_0$.
In the limit of vanishing photon field $A_\al$,
the explicit form of $\Qhat$ for arbitrary $d$
has been presented and studied in Ref.\ \cite{km13}.
The analogous Lagrange density 
for the quadratic part of the pure-photon sector at arbitrary $d$ 
is the subject of Ref.\ \cite{km09}.
Similar treatments exist for the nonminimal neutrino 
\cite{km12}
and gravity sectors 
\cite{nonmingrav}.

In the present work,
our focus is on operators having mass dimensions $d\leq 6$,
which are expected to generate the dominant physical effects
beyond the minimal SME.
The Lagrange density \rf{fermlag}
can be decomposed as the sum of the usual Dirac Lagrange density $\cl_0$
and a series of terms $\cl^{(d)}$ 
arising from the expansion of $\Qhat$ in operators of mass dimension $d$,
\beq
\cl_\ps =
\cl_0 
+ \cl^{(3)} 
+ \cl^{(4)} 
+ \cl^{(5)} 
+ \cl^{(6)} 
+ \ldots .
\label{nrlag}
\eeq
The explicit forms of the terms $\cl^{(3)}$ and $\cl^{(4)}$ 
are given in the original papers constructing the minimal SME
\cite{ck}
and are reproduced here for convenience, 
\bea
\cl^{(3)} &=&
- \acm 3 \mu \psb \ga_\mu \ps
- \bcm 3 \mu \psb \ga_5 \ga_\mu \ps
- \half \Hcm 3 \mn \psb \si_\mn \ps, 
\label{lag3}
\eea
and
\bea
\cl^{(4)} &=&
\half \ccm 4 {\ma} \psb \ga_\mu iD_\al \ps
+ {\rm h.c.}
\nn\\
&&
+ \half \dcm 4 {\ma} \psb \ga_5 \ga_\mu iD_\al \ps
+ {\rm h.c.}
\nn\\
&&
+ \half \ecm 4 {\al} \psb iD_\al \ps
+ {\rm h.c.}
\nn\\
&&
+ \half i \fcm 4 {\al} \psb \ga_5 iD_\al \ps
+ {\rm h.c.}
\nn\\
&&
+ \quar \gcm 4 {\mna} \psb \si_\mn iD_\al \ps
+ {\rm h.c.}
\label{lag4}
\eea
These terms have been the subject of numerous investigations,
and experimental constraints have been placed
on many of the corresponding coefficients
in several sectors of the SME
\cite{tables}.

At $d=5$, 
two kinds of terms enter the Lagrange density $\cl^{(5)}$,
one involving only symmetrized covariant derivatives $D_\al$ 
and one involving the electromagnetic field strength $F_\ab$,
\beq
\cl^{(5)} = \cl^{(5)}_D +\cl^{(5)}_F.
\eeq
The former is given explicitly by 
\bea
\cl^{(5)}_D &=&
- \half \mc 5 \ab \psb iD_{(\al} iD_{\be)} \ps
+ {\rm h.c.}
\nn\\
&&
- \half i \mfc 5 \ab \psb \ga_5 iD_{(\al} iD_{\be)} \ps
+ {\rm h.c.}
\nn\\
&&
- \half \ac 5 \mab \psb \ga_\mu iD_{(\al} iD_{\be)} \ps
+ {\rm h.c.}
\nn\\
&&
- \half \bc 5 \mab \psb \ga_5 \ga_\mu iD_{(\al} iD_{\be)} \ps
+ {\rm h.c.}
\nn\\
&&
- \quar \Hc 5 \mnab \psb \si_\mn iD_{(\al} iD_{\be)} \ps
+ {\rm h.c.}
\label{l5d}
\eea
The remaining piece is
\bea
\cl^{(5)}_F &=&
- \half \mcf 5 \ab F_\ab \psb \ps
- \half i \mfcf 5 \ab F_\ab \psb \ga_5 \ps
\nn\\
&&
- \half \acf 5 \mab F_\ab \psb \ga_\mu \ps
- \half \bcf 5 \mab F_\ab \psb \ga_5 \ga_\mu \ps
\nn\\
&&
- \quar \Hcf 5 \mnab F_\ab \psb \si_\mn \ps.
\label{l5f}
\eea
The two pieces $\cl^{(5)}_D$ and $\cl^{(5)}_F$ 
can be constructed as the symmetric and antisymmetric
combinations involving two covariant derivatives
because the electromagnetic field strength $F_\ab$
is obtained by commutation of covariant derivatives,
\beq
[i D_\al, i D_\be ] = - i q F_\ab.
\eeq
Within each piece $\cl^{(5)}_D$ and $\cl^{(5)}_F$,
the convenient separation of terms 
displayed in Eqs.\ \rf{l5d} and \rf{l5f}
reflects the decomposition of a $4\times 4$ spinor matrix
using the standard 16-component gamma-matrix basis.

For the Lagrange density at $d=6$,
three types of terms appear,
\beq
\cl^{(6)} = \cl^{(6)}_D + \cl^{(6)}_F + \cl^{(6)}_{\prt F}.
\eeq
The first involves only totally symmetrized combinations 
of three covariant derivatives,
\bea
\cl^{(6)}_D &=&
\half \cc 6 {\mabc} 
\psb \ga_\mu iD_{(\al} iD_\be iD_{\ga)} \ps 
+ {\rm h.c.}
\nn\\
&&
+ \half \dc 6 {\mabc} 
\psb \ga_5 \ga_\mu iD_{(\al} iD_\be iD_{\ga)} \ps 
+ {\rm h.c.}
\nn\\
&&
+ \half \ec 6 {\abc} 
\psb iD_{(\al} iD_\be iD_{\ga)} \ps 
+ {\rm h.c.}
\nn\\
&&
+ \half i \fc 6 {\abc} 
\psb \ga_5 iD_{(\al} iD_\be iD_{\ga)} \ps 
+ {\rm h.c.}
\nn\\
&&
+ \quar \gc 6 {\mnabc} 
\psb \si_\mn iD_{(\al} iD_\be iD_{\ga)} \ps 
+ {\rm h.c.}
\qquad
\label{l6d}
\eea
The second involves the field strength $F_{\al\be}$,
\bea
\cl^{(6)}_F &=&
\quar \ccf 6 {\mabc} F_{\bec} 
\big(
\psb \ga_\mu iD_{\al} \ps 
+ {\rm h.c.}
\big)
\nn\\
&&
+ \quar \dcf 6 {\mabc} F_{\bec} 
\big(
\psb \ga_5 \ga_\mu iD_{\al} \ps 
+ {\rm h.c.}
\big)
\nn\\
&&
+ \quar \ecf 6 {\abc} F_{\bec} 
\big(
\psb iD_{\al} \ps 
+ {\rm h.c.}
\big)
\nn\\
&&
+ \quar i \fcf 6 {\abc} F_{\bec} 
\big(
\psb \ga_5 iD_{\al} \ps 
+ {\rm h.c.}
\big)
\nn\\
&&
+ \frac 1 8 \gcf 6 {\mnabc} F_{\bec} 
\big(
\psb \si_\mn iD_{\al} \ps 
+ {\rm h.c.}
\big).
\label{l6f}
\eea
The remaining contributions involve the derivative
$\prt_\al F_{\be\ga}$ of the field strength, 
and they take the form
\bea
\cl^{(6)}_{\prt F} &=&
- \half \mcpf 6 \abc \prt_\al F_\bec ~\psb \ps
\nn\\
&&
- \half i \mfcpf 6 \abc \prt_\al F_\bec ~\psb \ga_5 \ps
\nn\\
&&
- \half \acpf 6 \mabc \prt_\al F_\bec ~\psb \ga_\mu \ps
\nn\\
&&
- \half \bcpf 6 \mabc \prt_\al F_\bec ~\psb \ga_5 \ga_\mu \ps
\nn\\
&&
- \quar \Hcpf 6 \mnabc \prt_\al F_\bec ~\psb \si_\mn \ps.
\label{l6df}
\eea

In constructing the above contributions to the Lagrange density $\cl_\ps$,
all of which are U(1) gauge invariant,
the coefficients for Lorentz violation
are assumed to be real
and can be taken as constant
in an inertial frame in the vicinity of the Earth
\cite{ck,akgrav}.
The dimension superscript $(d)$ 
is suppressed on minimal-SME coefficients.
Coefficients with subscript $F$ or $\prt F$ 
are associated with interactions directly involving 
the electromagnetic field strength or its derivative,
and they can be present even if the particle has zero charge.
The notation is chosen so that the indices $\mu$, $\nu$
are associated with spin properties,
while $\al$, $\be$, $\ga$ 
are associated with covariant momenta including field strengths.
Parentheses on $n$ indices imply symmetrization
with a factor of $1/n!$.
The index symmetries of the coefficients
are otherwise evident by inspection.

\renewcommand{\arraystretch}{1.2}
\begin{table}
\caption{
\label{nrcoeff}
Properties of terms in $\cl^{(d)}$ for $d\leq 6$.} 
\setlength{\tabcolsep}{5pt}
\begin{tabular}{cccc}
\hline
\hline
$d$ & Coefficient & CPT & Number \\
\hline
3	&	$	 \acm 3 \mu	$	&	odd	&	4	\\	
	&	$	\bcm 3 \mu	$	&	odd	&	4	\\	
	&	$	\Hcm 3 \mn	$	&	even	&	6	\\	[5pt]
4	&	$	 \ccm 4 {\ma}	$	&	even	&	10	\\	
	&	$	\dcm 4 {\ma}	$	&	even	&	10	\\	
	&	$	\ecm 4 {\al}	$	&	odd	&	4	\\	
	&	$	\fcm 4 {\al}	$	&	odd	&	4	\\	
	&	$	\gcm 4 {\mna}	$	&	odd	&	24	\\	[5pt]
5	&	$	\mc 5 \ab	$	&	even	&	10	\\	
	&	$	\mfc 5 \ab	$	&	even	&	10	\\	
	&	$	\ac 5 \mab	$	&	odd	&	40	\\	
	&	$	\bc 5 \mab	$	&	odd	&	40	\\	
	&	$	 \Hc 5 \mnab	$	&	even	&	60	\\	[3pt]
	&	$	 \mcf 5 \ab	$	&	even	&	6	\\	
	&	$	\mfcf 5 \ab	$	&	even	&	6	\\	
	&	$	\acf 5 \mab	$	&	odd	&	24	\\	
	&	$	\bcf 5 \mab	$	&	odd	&	24	\\	
	&	$	\Hcf 5 \mnab	$	&	even	&	36	\\	[5pt]
6	&	$	\cc 6 {\mabc}	$	&	even	&	80	\\	
	&	$	\dc 6 {\mabc}	$	&	even	&	80	\\	
	&	$	\ec 6 {\abc}	$	&	odd	&	20	\\	
	&	$	\fc 6 {\abc}	$	&	odd	&	20	\\	
	&	$	\gc 6 {\mnabc}	$	&	odd	&	120	\\	[3pt]
	&	$	\ccf 6 {\mabc}	$	&	even	&	96	\\	
	&	$	\dcf 6 {\mabc}	$	&	even	&	96	\\	
	&	$	 \ecf 6 {\abc}	$	&	odd	&	24	\\	
	&	$	 \fcf 6 {\abc}	$	&	odd	&	24	\\	
	&	$	\gcf 6 {\mnabc}	$	&	odd	&	144	\\	[3pt]
	&	$	 \mcpf 6 \abc	$	&	odd	&	20	\\	
	&	$	\mfcpf 6 \abc	$	&	odd	&	20	\\	
	&	$	\acpf 6 \mabc	$	&	even	&	80	\\	
	&	$	\bcpf 6 \mabc	$	&	even	&	80	\\	
	&	$	\Hcpf 6 \mnabc	$	&	odd	&	120	\\	
\hline
\hline
\end{tabular}
\end{table}

Table \ref{nrcoeff} lists some properties of the terms 
appearing in the expansion of the Lagrange density \rf{nrlag}
for $d\leq 6$.
The first column gives the dimension of the Lorentz-violating operator, 
while the second lists the corresponding coefficient.
The units of each coefficient are GeV$^{4-d}$.
The CPT parity of the operator is presented in the third column.
The final column displays
the number of independent operators.
Note that this counting incorporates the constraints
from the Bianchi identity,
which here is equivalent to the usual homogeneous Maxwell equations  
$\ep^\abcd \prt_\ga F_\ab = 0$.

In the limit of vanishing $A_\al$, 
the contributions \rf{l5d} and \rf{l6d}
are the leading-order nonminimal terms
in the general treatment of Dirac fermions 
in the presence of Lorentz-violating operators at arbitrary $d$,
which includes various special models as limiting cases
\cite{km13}.
Experimental constraints on some of the corresponding coefficients
in the electron, proton, and muon sectors have been obtained 
\cite{tables}.
The same coefficients control all terms in $\cl^{(5)}_D$ and $\cl^{(6)}_D$,
even when $A_\al$ is nonzero,
so the corresponding constraints hold 
in any models built from these terms as well.

In contrast,
the literature lacks
a systematic treatment of $d\leq 6$ Lorentz-violating spinor couplings 
to the field strength $F_\ab$ and its derivative $\prt_\ga F_\ab$.
The set of operators involving these couplings 
in the above expressions for
$\cl^{(5)}_F$, $\cl^{(6)}_F$, and $\cl^{(6)}_{\prt F}$
provides a complete enumeration for $d\leq 6$
and encompasses all possible models of this type.
A subset of these terms appears naturally in noncommutative QED
\cite{chklo01},
where the noncommutativity parameter $\th^{\ab}\equiv -i[x^\al, x^\be]$
generates coefficients for Lorentz violation at $d=5$ and $d=6$
according to 
\bea
\mfcf 5 \ab &\to& 
-\half m q \th^{\ab},
\nn\\
\ccf 6 {\mabc} &\to& 
-\half q (\et^{\mu\al}\th^{\be\ga}+2\et^{\mu[\be}\th^{\ga]\al}).
\eea
The work of Belich, Costa-Soares, Ferreira, and Helay\"el-Neto
\cite{bcfh05},
which studies the special Lorentz-violating limits 
\bea
\acf 5 \mab \to g\ep^{\mab}{}_{\ga}v^{\ga},
\quad
\bcf 5 \mab \to -g_{a}\ep^{\mab}{}_{\ga}v^{\ga},
\eea
spawned numerous followup investigations  
of models restricted to specifically chosen 
Lorentz-violating operators with spinor couplings to $F_\ab$
\cite{Fmodels}.
Also,
a model containing all $d=5$ operators that
cannot be reduced via equations of motion to ones with $d<5$  
has been given in Ref.\ \cite{bo08},
using a different organization of terms than adopted here. 

Actual constraints on physical effects from 
$\cl^{(5)}_F$, $\cl^{(6)}_F$, and $\cl^{(6)}_{\prt F}$
have so far been obtained 
on only a small part of the available coefficient space 
displayed in Table \ref{nrcoeff}
\cite{chklo01,ar15,ar16,edm}.
For anomalous magnetic moments,
which are the focus of the sections that follow,
the sole limits to date have been reported recently
by Araujo, Casana, and Ferreira
\cite{ar15,ar16},
who consider in turn several special Lorentz-violating limits 
of the coefficient $\Hcf 5 \mnab$ given by
\bea
\Hcf 5 \mnab &\to& -2\la(K_F)^{\mn\ab},
\nn\\
\Hcf 5 \mnab&\to&-\la_A\ep_{\rh\si}{}^{\mn}(K_F)^{\rh\si\ab},
\nn\\
\Hcf 5 \mnab &\to&
-2\la_1^{\prime}(\et^{\al[\mu}T^{\nu]\be}-\et^{\be[\mu}T^{\nu]\al}),
\nn\\
\Hcf 5 \mnab &\to& 
\frac 34\la_3(\et^{\mu[\al}T^{\be]\nu}-\et^{\nu[\al}T^{\be]\mu}),
\eea
where $(K_F)^{\rh\si\ab}$ is taken to 
have the symmetries of the Riemann tensor.

The above discussions of both the theoretical and experimental implications
of terms with spinor couplings to $F_\ab$
are further convoluted by the possibility of removing 
the corresponding Lorentz-violating operators 
from the Lagrange density using field redefinitions.
We show in Sec.\ \ref{Field redefinitions} below 
that this possibility,
which has been overlooked in the literature to date, 
implies that only certain combinations of these terms
can produce observable effects in experiments.
Remarkably,
it turns out that many specific spinor couplings to $F_\ab$ 
at finite $d$
can be removed from observables in favor of other terms,
including in particular the coefficient $\Hcf 5 \mnab$.

We emphasize that the Lagrange density \rf{nrlag}
also contains all Lorentz-{\it invariant} fermion-photon couplings.
These terms arise from components of the SME coefficients
that are proportional to the Minkowski-metric or Levi-Civita tensors, 
both of which are Lorentz invariant.
The only Lorentz-invariant terms arising in the minimal SME are
\beq
\cl^{(4)}_{\rm LI} =
\half \cc 4 {} \psb \ga^\mu iD_\mu \ps
+ \half \dc 4 {} \psb \ga_5 \ga^\mu iD_\mu \ps
+ {\rm h.c.}
\label{lilag4}
\eeq
These terms can be absorbed into the normalizations 
of the left- and right-handed components of the spinor field $\ps$.
Both are typically assumed to vanish
in the literature on Lorentz violation.

The Lorentz-invariant terms of mass dimension $d=5$
can be written explicitly as
\bea
\cl^{(5)}_{\rm LI} &=&
- \half \mc {5} {} \psb (iD)^2 \ps
+ {\rm h.c.}
\nn\\
&&
- \half i \mfc {5} {} \psb \ga_5 (iD)^2\ps
+ {\rm h.c.}
\nn\\
&&
- H^{(5)}_{F,1} F^\mn \psb \si_\mn \ps
- H^{(5)}_{F,2} {\widetilde F}^\mn \psb \si_\mn \ps,
\label{lilag5}
\eea
where ${\widetilde F}^\mn = \ep^{\mn\ab} F_\ab/2$.
These terms include Lorentz-invariant contributions 
to the anomalous magnetic and electric moments of the spinor field $\ps$
involving the coefficients $H^{(5)}_{F,1}$ and $H^{(5)}_{F,2}$.
Finally,
the Lorentz-invariant terms with $d=6$ are
\bea
\cl^{(6)}_{\rm LI} &=&
+\frac 1 6 \cc 6 {} 
\psb \ga^\mu [iD_{\mu} (iD)^2 + iD^\al iD_{\mu} iD_\al 
\nn\\
&&
\hskip 50pt
+ (iD)^2 iD_{\mu}] \ps 
+ {\rm h.c.}
\nn\\
&&
+ \half c^{(6)}_{F,1} F^\mn 
(\psb \ga_\mu iD_\nu \ps 
+ {\rm h.c.})
\nn\\
&&
+ \half c^{(6)}_{F,2} {\widetilde F}^\mn 
(\psb \ga_\mu iD_\nu \ps + {\rm h.c.})
\nn\\
&&
+\frac 1 6 \dc 6 {} 
\psb \ga_5\ga^\mu [iD_{\mu} (iD)^2 + iD^\al iD_{\mu} iD_\al 
\nn\\
&&
\hskip 50pt
+ (iD)^2 iD_{\mu}] \ps 
+ {\rm h.c.}
\nn\\
&&
+ \half d^{(6)}_{F,1} F^\mn 
(\psb \ga_5 \ga_\mu iD_\nu \ps + {\rm h.c.})
\nn\\
&&
+ \half d^{(6)}_{F,2} {\widetilde F}^\mn 
(\psb \ga_5 \ga_\mu iD_\nu \ps 
+ {\rm h.c.})
\nn\\
&&
- a^{(6)}_{\prt F, 1} \prt_\al F^\ab \psb \ga_\be \ps
- a^{(6)}_{\prt F, 2} \prt_\al {\widetilde F}^\ab \psb \ga_\be \ps
\nn\\
&&
- b^{(6)}_{\prt F, 1} \prt_\al F^\ab \psb \ga_5 \ga_\be \ps
- b^{(6)}_{\prt F, 2} \prt_\al {\widetilde F}^\ab \psb \ga_5 \ga_\be \ps,
\nn\\
\label{lilag6}
\eea
where the homogeneous Maxwell equations have been used.

Note that the possibility of using field redefinitions
to remove some terms in the Lagrange density in favor of others 
also applies to the Lorentz-invariant operators in
$\cl^{(5)}_{\rm LI}$ and $\cl^{(6)}_{\rm LI}$.
One example in the next subsection illustrates this 
by absorbing the conventional couplings 
$H^{(5)}_{F,1}$ and $H^{(5)}_{F,2}$
for the anomalous magnetic and electric moments
into other coefficients.

\subsection{Field redefinitions}
\label{Field redefinitions}

The freedom to choose canonical dynamical variables 
via suitable field redefinitions
often implies that two seemingly different theories
in fact describe the same physics.
For example,
in the context of the standard kinetic term for a Dirac fermion,
a chiral rotation of the field $\ps$ can absorb
a possible term $- i \mfcm 3 {} \psb \ga_5 \ps$ 
into the usual mass term modulo anomaly considerations,
leaving $\m$ as the fermion mass.

In the context of the SME,
field redefinitions reveal that some terms 
that naively appear to violate Lorentz symmetry
have no measurable implications,
while others are observable only in certain specific combinations
\cite{ck,akgrav,km13,redefref}.
The simplest example involving Lorentz and CPT violation
is a linear phase redefinition
of the form $\ps = \exp(-i a^\mu x_\mu) \ch$,
which physically redefines the zero of energy and momentum
and can be used to eliminate 
the term $\acm 3 \mu \psb \ga_\mu \ps$
from $\cl$ at leading order in Lorentz violation. 
We remark in passing that the contribution from $a_\mu$ is distinct 
from that due to a constant 4-potential $A_\mu$
because $a_\mu$ is gauge invariant
and so cannot be removed by a gauge transformation.

In this subsection,
we examine the effects of certain field redefinitions
on the terms in $\cl_\ps$ with $d\leq 6$.
Specific results are extracted for a constant electromagnetic field,
which is the scenario of relevance for many experimental applications.

\subsubsection{Fermion redefinitions}
\label{Fermion redefinitions}

We consider here gauge-covariant field redefinitions
amounting to renormalizations of $\ps$ 
taking the form
\beq
\ps = (1+\Zhat) \ps' ,
\label{redef}
\eeq
where we allow $\Zhat$ to depend on covariant derivatives.
Under this transformation,
the physics is invariant provided
the Lorentz-violating terms in $\cl$ remain perturbative,
which holds if $\Zhat$ itself is perturbative
\cite{km13}. 
Note that this implies both the field strength $F_\ab$ 
and the coefficients for Lorentz violation must be small
on the scale of the energies and momenta of interest.

For notational simplicity,
it is convenient to work in momentum space,
writing $p_\al = i D_\al$ and
\beq
[p_\al, p_\be] = -iq F_\ab.
\label{pcomm}
\eeq
The redefinition \rf{redef} induces a new operator $\Qhat'$ 
from the Lagrange density \rf{nrlag},
\beq
\ps^\dag \ga_0(\ga^\mu p_\mu - \m + \Qhat) \ps
\approx
\ps'^\dag \ga_0(\ga^\mu p_\mu - \m + \Qhat') \ps' ,
\eeq
where
\beq
\Qhat' = \Qhat 
+ (\ga^\mu p_\mu - \m) \Zhat + \ga_0\Zhat^\dag\ga_0 (\ga^\mu p_\mu - \m) .
\eeq
For convenience,
we can separate $\Zhat$ 
into a hermitian piece $\Xhat$ and an antihermitian piece $\Yhat$
given by
\beq
\Zhat = \Xhat+i\Yhat , 
\hskip 4pt
\Xhat = \half(\Zhat+\ga_0\Zhat^\dag\ga_0) , 
\hskip 4pt
\Yhat = \tfrac{1}{2i}(\Zhat-\ga_0\Zhat^\dag\ga_0) .
\eeq
Note that $\Xhat$ and $\Yhat$ satisfy 
the hermiticity conditions
$\Xhat = \ga_0 \Xhat^\dag\ga_0$,
$\Yhat = \ga_0 \Yhat^\dag\ga_0$,
in parallel with the hermiticity of the $\Qhat$ operator.
Using these definitions,
the shift $\de\Qhat = \Qhat' - \Qhat$ 
in the modified Dirac operator $\Qhat$
arising from the field redefinition is found to be
\bea
\de\Qhat 
&=&
-2\m\Xhat 
+ \{p_\mu \ga^\mu,\Xhat\} + i [ p_\mu \ga^\mu,\Yhat] 
\nn\\
&=& 
-2\m\Xhat + p_\mu \{\ga^\mu,\Xhat\} + i p_\mu[\ga^\mu,\Yhat] 
\nn\\
&&
- [p_\mu ,\Xhat] \ga^\mu + i [p_\mu ,\Yhat] \ga^\mu .
\label{delQ}
\eea

Explicit expressions for the shift $\de\Qhat$
can be found via decomposition of $\Xhat$ and $\Yhat$
in terms of the basis of 16 Dirac matrices
and power series in $p_\al$.
First,
we define
\bea
\Xhat 
&=& \Xhat_I\Ga^I
\nn\\
&\equiv& 
\Xhat_S
+ i\Xhat_P\ga_5
+ \Xhat_V^\mu \ga_\mu
+ \Xhat_A^\mu \ga_5\ga_\mu
+ \half \Xhat_T^\mn \si_\mn ,
\nn\\
\Yhat 
&=& \Yhat_I\Ga^I
\nn\\
&\equiv& \Yhat_S
+ i\Yhat_P \ga_5
+ \Yhat_V^\mu \ga_\mu
+ \Yhat_A^\mu \ga_5\ga_\mu
+ \half \Yhat_T^\mn \si_\mn .
\qquad
\label{XYexp}
\eea
Here, 
the index $I$ takes values $S$, $P$, $V$, $A$, $T$
and is summed.
Each component in these expressions
can then in turn be expanded in powers of $p_\al$,
\bea
\Xhat_I^\vs
&=&
X_I^\vs + X_I^{\va} p_\al + X_I^\vab p_\al p_\be 
+ X_I^\vabc p_\al p_\be p_\ga + \ldots,
\nn\\
\Yhat_I^\vs 
&=&
Y_I^\vs + Y_I^{\va} p_\al + Y_I^{\vab} p_\al p_\be 
+Y _I^{\vabc} p_\al p_\be p_\ga + \ldots,
\nn\\
\label{momedec}
\eea
where the index $\vs$ takes values 
that are null, $\mu$, or $\mn$
according to the Lorentz properties 
of the corresponding spinor matrix.
Note that the ordering of the momenta in this expression is significant
because they have nonzero commutators.
Via this procedure, 
all the spin and momentum dependence is explicitly extracted
and so the components appearing in the decomposition \rf{momedec}
are merely constants.

In studying the possible shifts $\de\Qhat$ induced by field redefinitions,
each of the constant components can be treated 
as inducing an independent field redefinition.
Each of these is U(1) gauge covariant by construction.
It suffices for present purposes to keep terms up to third order in $p_\al$.
Since there are two pieces $\Xhat$, $\Yhat$,
each of which has five spin components,
each of which has four momentum components,
we see that the above decomposition
allows for 40 distinct field redefinitions
in this language. 
Note that some redefinitions duplicate effects
and some redefinitions induce multiple coefficient shifts.
The redefinition $X_S$ introduces an irrelevant scaling
of the usual Dirac action,
while the redefinition $Y_S$ has no effect.

As a simple example,
consider the field redefinition associated with 
$\Yhat \supset Y_S^\al p_\al$.
The result \rf{delQ} implies 
\bea
\de\Qhat 
= - q Y_S^{[\al}\et^{\be]\mu} F_{\al\be} \ga_\mu
\leftrightarrow 
- \half \acf 5 \mab F_\ab \ga_\mu ,
\eea
where the brackets around index pairs 
indicate antisymmetrization with a factor of 1/2.
The last part of this expression gives the match 
to the corresponding term in the Lagrange density \rf{l5f}.
The shift $\de \acf 5 \mab$ induced 
in the coefficient $\acf 5 \mab$ 
via this field redefinition is therefore  
$\de \acf 5 \mab = 2 q Y_S^{[\al}\et^{\be]\mu}$.
One consequence of this result 
is that the trace of the mixed-symmetry representation
in $\acf 5 \mab$ has no independent physical content
and hence cannot be measured independently in experiments.

As a more involved example,
consider the redefinition associated with 
$\Xhat \supset X_V^\ma p_\al \ga_\mu$.
We obtain
\bea
\de\Qhat &=&
- 2 \m X_V^\ma p_\al \ga_\mu
+ 2 X_V^\ma p_{(\mu} p_{\al)}
\nn\\
&&
\quad
- \half q( X_V^{\mu[\al} \et^{\be]\nu} 
- X_V^{\nu[\al} \et^{\be]\mu}) 
F_{\al\be} \si_{\mu\nu} .
\eea
The correspondence 
to terms in the Lagrange density \rf{nrlag} yields 
\bea
\de \ccm 4 {\ma} &=& - 2\m X_V^\ma ,
\nn\\
\de \mc 5 \ab 
&=&
- 2 X_V^{(\ab)} ,
\nn\\
\de \Hcf 5 \mnab 
&=&
2q (X_V^{\mu[\al} \et^{\be]\nu} - X_V^{\nu[\al} \et^{\be]\mu}) .
\eea
The parameter $X_V^\ma$ can itself be decomposed
into symmetric traceless, antisymmetric, and trace pieces,
each of which can also be viewed as an independent redefinition.
The above equations therefore reproduce the known result
that the antisymmetric part of $\ccm 4 {\ma}$ is unphysical
\cite{ck}
and reveal that the coefficient $\mc 5 \ab$ can be removed
by absorption into $X_V^{\mu[\al} \et^{\be]\nu}$.

We provide here one final explicit example,
based on the redefinition associated with
$\Xhat\supset \half \si_{\mn} X_T^{\mn\ab}p_{\al}p_{\be}$.
Some calculation yields
\bea
\de \Qhat
&=&
- \m X_T^{\mn(\ab)}\si_{\mn} p_{(\al} p_{\be)}
\nn\\
&& 
+ \half i \m q X_T^{\mn[\ab]} F_{\ab} \si_{\mn}
\nn\\
&&
+ X_T^{\rh\si(\ab} \ep^{\ga)\mu}{}_{\rh\si} 
p_{(\al} p_{\be} p_{\ga)} \ga_5\ga_{\mu}
\nn\\
&&
+ q (X_T^{\be\mu(\al\ga)} - X_T^{\ga\mu(\ab)})
F_{\be\ga} p_{\al} \ga_{\mu}
\nn\\
&&
- \half i q X_T^{\rh\si[\be\ga]} \ep^{\al\mu}{}_{\rh\si}
F_{\be\ga} p_{\al} \ga_5\ga_{\mu}.
\eea
This generates coefficient shifts given by
\bea
\de \Hc 5 \mnab &=& 2 \m X_T^{\mn(\ab)},
\nn\\
\de \Hcf 5 \mnab &=& -2 i \m q X_T^{\mn[\ab]},
\nn\\
\de \dc 6 {\mabc} &=& X_T^{\rh\si (\ab} \ep^{\ga)\mu}{}_{\rh\si},
\nn\\
\de \ccf 6 {\mabc} &=& 2 q (X_T^{\be\mu(\al\ga)} - X_T^{\ga\mu(\ab)}),
\nn\\
\de \dcf 6 {\mabc} &=& i q X_T^{\rh\si[\be\ga]} \ep^{\mu\al}{}_{\rh\si}.
\label{xtredef}
\eea
Among the implications of these equations
is that the coefficient $\Hcf 5 \mnab$,
which controls $d=5$ spinor couplings to $F_\ab$
and has been a popular subject of investigation in the literature,
can be absorbed into the coefficient $\dcf 6 {\mabc}$.
This point is discussed further in a more general context
in the following subsection.

\renewcommand{\arraystretch}{1.0}
\begin{table}
\caption{
\label{fieldredefs}
Effects of field redefinitions for $d\leq 6$.} 
\setlength{\tabcolsep}{2pt}
\begin{tabular}{ccl}
\hline
\hline
$d$ & Shift & Field redefinition \\
\hline
3	&	$	\de	 \acm 3 \mu	$	&	$	2\m X_V^\mu	$	\\	
	&	$	\de	\bcm 3 \mu	$	&	$	2\m X_A^\mu	$	\\	
	&	$	\de	\Hcm 3 \mn	$	&	$	2\m X_T^{\mn}	$	\\	[5pt]
4	&	$	\de	 \ccm 4 {\ma}	$	&	$	2X_S\et^{\ma}, -2\m X_V^{\ma}	$	\\	
	&	$	\de	\dcm 4 {\ma}	$	&	$	X_T^{\nu\rh}\ep^{\ma}{}_{\nu\rh}, 2Y_P\et^{\ma}, -iY_T^{\nu\rh}\ep^{\mu\al}{}_{\nu\rh}	$	\\	
	&	$	\de	\ecm 4 {\al}	$	&	$	-2\m X_S^\al, 2X_V^\al	$	\\	
	&	$	\de	\fcm 4 {\al}	$	&	$	-2\m X_P^\al	$	\\	
	&	$	\de	\gcm 4 {\mna}	$	&	$	2X_A^\rh\ep_\rh{}^{\mna}, 2\m X_T^{[\mn]\al},	$	\\	
	&	$			$	&	$	 -4Y_V^{[\mu}\et^{\nu]\al}, 2iY_A^\rh\ep_\rh{}^{\mn\al}	$	\\	[5pt]
5	&	$	\de	\mc 5 \ab	$	&	$	2\m X_S^{(\ab)}, -2X_V^{(\ab)}	$	\\	
	&	$	\de	\mfc 5 \ab	$	&	$	2\m X_P^{(\ab)}, 2Y_A^{(\ab)}	$	\\	
	&	$	\de	\ac 5 \mab	$	&	$	2X_S^{(\al}\et^{\be)\mu}, 2\m X_V^{\mu(\ab)}, 2Y_T^{\mu(\ab)}	$	\\	
	&	$	\de	\bc 5 \mab	$	&	$	2\m X_A^{\mu(\ab)}, X_T^{\nu\rh(\al}\ep^{\be)\mu}{}_{\nu\rh}, iY_P^{(\al}\et^{\be)\mu}	$	\\	
	&	$	\de	 \Hc 5 \mnab	$	&	$	2\m X_T^{\mn(\ab)}, 4Y_V^{[\mu|(\al}\et^{\be)|\nu]} 	$	\\	[3pt]
	&	$	\de	 \mcf 5 \ab	$	&	$	 -2i\m qX_S^{[\ab]}, -2qY_V^{[\ab]}	$	\\	
	&	$	\de	\mfcf 5 \ab	$	&	$	 -2i\m qX_P^{[\ab]}, -2qX_A^{[\ab]}	$	\\	
	&	$	\de	\acf 5 \mab	$	&	$	 -2iq\m X_V^{\mu[\ab]}, 2qX_T^{\mu[\ab]}, 2qY_S^{[\al}\et^{\be]\mu}	$	\\	
	&	$	\de	\bcf 5 \mab	$	&	$	 -2qX_P^{[\al}\et^{\be]\mu}, -2i\m qX_A^{\mu[\ab]}, qY_T^{\nu\rh[\al}\ep^{\be]\mu}{}_{\nu\rh}	$	\\	
	&	$	\de	\Hcf 5 \mnab	$	&	$	2q(X_V^{\mu[\al} \et^{\be]\nu}-X_V^{\nu[\al} \et^{\be]\mu}), 4X_A^{\rh[\al}\ep^{\be]\mn}{}_{\rh}, 	$	\\	
	&	$			$	&	$	  -2i\m qX_T^{\mn[\ab]}, -2qY_A^{\rh[\al}\ep^{\be]\mn}{}_{\rh}	$	\\	[5pt]
6	&	$	\de	\cc 6 {\mabc}	$	&	$	2X_S^{(\ab}\et^{\ga)\mu}, -2\m X_V^{\mu(\ab\ga)}, Y_T^{\mu(\ab\ga)}	$	\\	
	&	$	\de	\dc 6 {\mabc}	$	&	$	-2\m X_A^{\mu(\ab\ga)}, X_T^{\nu\rh(\ab}\ep^{\ga)\mu}{}_{\nu\rh}, -2iY_P^{(\ab}\et^{\ga)\mu}	$	\\	
	&	$	\de	\ec 6 {\abc}	$	&	$	-2\m X_S^{(\ab\ga)}, 4X_V^{(\abc)}	$	\\	
	&	$	\de	\fc 6 {\abc}	$	&	$	 -2\m X_P^{(\ab\ga)}, -2Y_A^{\ab\ga}	$	\\	
	&	$	\de	\gc 6 {\mnabc}	$	&	$	 -2X_A^{\rh(\ab}\ep^{\ga)\mn}{}_{\rh}, -2\m X_T^{\mn(\ab\ga)}	$	\\	
	&	$			$	&	$	-4Y_V^{[\mu|(\ab}\et^{\ga)|\nu]} 	$	\\	[3pt]
	&	$	\de	\ccf 6 {\mabc}	$	&	$	 -2iqX_S^{[\be\ga]}\et^{\al\mu}, 4i\m qX_V^{\mu\langle\al[\be\ga]\rangle}, 	$	\\	
	&	$			$	&	$	 2q(X_T^{\be\mu(\al\ga)}-X_T^{\ga\mu(\ab)}),	$	\\	
	&	$			$	&	$	 -2q(Y_S^{(\ab)}\et^{\ga\mu}-Y_S^{(\al\ga)}\et^{\be\mu}), -2iqY_T^{\mu\al[\be\ga]} 	$	\\	
	&	$	\de	\dcf 6 {\mabc}	$	&	$	 2q(X_P^{(\ab)}\et^{\ga\mu}-X_P^{(\al\ga)}\et^{\be\mu}), 4i\m qX_A^{\mu\langle\al[\be\ga]\rangle},	$	\\	
	&	$			$	&	$	  -iqX_T^{\nu\rh[\be\ga]}\ep^{\al\mu}{}_{\nu\rh}, 2qY_P^{[\be\ga]}\et^{\al\mu},	$	\\	
	&	$			$	&	$	 -q(Y_T^{\nu\rh(\ab)}\ep^{\ga\mu}{}_{\nu\rh}-Y_T^{\nu\rh(\al\ga)}\ep^{\be\mu}{}_{\nu\rh})	$	\\	
	&	$	\de	 \ecf 6 {\abc}	$	&	$	 4i\m qX_S^{\langle\al[\be\ga]\rangle}, -4iqX_V^{\al[\be\ga]},	$	\\	
	&	$			$	&	$	 2q(Y_V^{\be(\al\ga)}-Y_V^{\ga(\ab)})	$	\\	
	&	$	\de	 \fcf 6 {\abc}	$	&	$	 4i\m qX_P^{\langle\al[\be\ga]\rangle}, 2q(X_A^{\be(\al\ga)}-X_A^{\ga(\ab)}),	$	\\	
	&	$			$	&	$	  2iqY_A^{\al[\be\ga]}	$	\\	
	&	$	\de	\gcf 6 {\mnabc}	$	&	$	 -4q(X_V^{[\mu|(\ab)}\et^{\ga|\nu]}-X_V^{[\mu|(\al\ga)}\et^{\be|\nu]}),	$	\\	
	&	$			$	&	$	 2iqX_A^{\rh[\be\ga]}\ep^{\al\mn}{}_{\rh},  4i\m qX_T^{\mn\langle\al[\be\ga]\rangle}, 	$	\\	
	&	$			$	&	$	4iqY_V^{\mu[\be\ga]}\et^{\al\nu},	$	\\	
	&	$			$	&	$	 -2q(Y_A^{\rh(\ab)}\ep^{\ga\mn}{}_{\rh}-Y_A^{\rh(\al\ga)}\ep^{\be\mn}{}_{\rh})	$	\\	[5pt]
\hline
\hline
\end{tabular}
\end{table}

A related and striking observation is that
the standard Lorentz-invariant terms 
describing anomalous magnetic and electric dipole moments
can be removed from the Lagrange density
by using a special limit of the redefinition \rf{xtredef}.
Suppose a fermion is described by the conventional Dirac Lagrange density
plus the specific coupling 
in Eq.\ \rf{lilag5} involving $H^{(5)}_{F,1}$.
Choosing 
\beq
X_T^{\mn[\ab]} = -\fr{i}{\m q} H^{(5)}_{F,1}
(\et^{\mu[\al}\et^{\be]\nu} - \et^{\nu[\al}\et^{\be]\mu})
\eeq
and performing the corresponding redefinition 
removes the operator $F^\mn \psb \si_\mn \ps$
with coupling $H^{(5)}_{F,1}$
in favor of the operator with coupling $d^{(6)}_{F,2}$
in the Lagrange density \rf{lilag6}.
With a similar redefinition,
the coupling $H^{(5)}_{F,2}$ in $\cl^{(5)}_{\rm LI}$ 
can be absorbed into the coupling $d^{(6)}_{F,1}$ in $\cl^{(6)}_{\rm LI}$.
Explicitly,
these redefinitions implement the transformations
\bea
(H^{(5)}_{F,1}, d^{(6)}_{F,2}\equiv 0)
&\to &
(H^{(5)}_{F,1}\equiv 0, d^{(6)}_{F,2} = 2 H^{(5)}_{F,1}/\m),
\nn\\
(H^{(5)}_{F,2}, d^{(6)}_{F,1}\equiv 0)
&\to &
(H^{(5)}_{F,2}\equiv 0, d^{(6)}_{F,1} = -2 H^{(5)}_{F,2}/\m),
\nn\\
\eea 
thereby moving the usual $d=5$ Lorentz-invariant 
anomalous magnetic and electric dipole moments
to the $d=6$ nonminimal Lorentz-invariant sector. 

For the general case,
the results of each field redefinition considered in turn
are displayed in Table \ref{fieldredefs}.
The first column gives the dimension of the operator.
The second column shows the coefficient shift being considered.
The third column indicates the structure of the field redefinition
implementing the coefficient shift.
Most coefficients are shifted
by more than one field redefinition,
and the different redefinitions are separated by commas.
Notable exceptions are the coefficients appearing in 
$\cl^{(6)}_{\prt F}$,
for which the corresponding operators play no role 
in the field redefinitions due to the Bianchi identity
and which are therefore omitted from the table.
As before,
parentheses and brackets 
around $n$ indices imply symmetrization
and antisymmetrization,
respectively,
with a factor of $1/n!$ included.
A few terms involve the specific index combination
of three indices that we denote by chevrons,
\beq
\vev{\abc}\equiv \half (\al\be\ga + \be\ga\al - \ga\al\be).
\eeq
Following standard convention,
vertical bars are used to denote indices
omitted from the symmetrization or antisymmetrization.

\subsubsection{Absorption of couplings to $F_\ab$}
\label{Absorption}

As an interesting and potentially useful example 
of the application of field redefinitions,
we investigate in this subsection
the possibility of absorbing a given spinor coupling to $F_\ab$
into other terms in the Lagrange density.
The discussion considers operators of any mass dimension $d$.

The terms of primary interest for this illustrative calculation
are the gauge-invariant spinor couplings to the field strength $F_\ab$,
where $F_\ab$ may be nonconstant.
For definiteness,
we focus here on terms involving exactly one power of $F_\ab$,
rather than those nonlinear in $F_\ab$ 
or those involving derivatives of $F_\ab$. 
In momentum space,
the corresponding operators 
can be collected in a quantity $\Qhat_F$ taking the form
\beq
\Qhat_F = 
\sum_{d>4}
k_I^{(d)\vs\al\be\al_1 \ldots \al_{d-5}} 
F_\ab ~p_{(\al_1} \ldots p_{\al_{d-5})}
\Ga^I_\vs,
\label{qf}
\eeq 
where 
$k_I^{\vs\al\be\al_1 \ldots \al_j}$ 
are $F$-type coefficients for Lorentz violation.
As before,
the index $I$ takes values $S$, $P$, $V$, $A$, $T$,
while $\vs$ is null or takes values $\mu$ or $\mn$.
For example,
the operators appearing in 
the expression \rf{l5f} for $\cl^{(5)}_F$ 
and the expression \rf{l6f} for $\cl^{(6)}_F$
are reproduced by particular terms in the series \rf{qf}. 

For definiteness,
we examine here the term $-2\m \Xhat$ 
in the shift \rf{delQ} of $\de \Qhat$
and investigate its implications for terms in the series \rf{qf}.
The quantity $\Xhat$ can be expanded 
according to the expression \rf{XYexp},
and then each resulting piece $\Xhat_I^\vs$ 
can itself be expanded in covariant momenta 
following the definition \rf{momedec}.
A given term in the expansion \rf{momedec}
of $\Xhat_I^\vs$ 
can be viewed as a sum of irreducible representations
obtained by decomposing the product of momenta.
The commutator \rf{pcomm} of any two momenta
produces a factor of $F_\ab$,
so the expansion \rf{momedec} can be seen 
to contain a series of terms involving powers of $F_\ab$.
Here,
the interest lies in terms with a single factor of $F_\ab$,
corresponding to the representation
with all momenta symmetrized except for a single pair.
Denoting this representation by $\{\al\be\ldots\}$,
the expansion \rf{momedec} contains
\bea
\Xhat_I^\vs
&\supset& 
- \half i q ( X_I^\vab F_\ab 
+ X_I^\vabc F_{\{\ab} p_{\ga\}} 
\nn\\
&& 
\hskip 30pt
+ X_I^\vabcd F_{\{\ab} p_\ga p_{\de\}} 
+ \ldots)
\nn\\
&=& 
- \half i q ( X_I^\vab F_\ab 
+ X_I^{\vs\{\abc\}} F_{\ab} p_{\ga} 
\nn\\
&& 
\hskip 30pt
+ X_I^{\vs\{\abcd\}} F_{\ab} p_{(\ga} p_{\de)} 
+ \ldots).
\eea
It follows that
\beq
\Xhat 
\supset
- \half i q 
\sum_{d>4}
X_I^{\vs\{\al\be\al_1 \ldots \al_{d-5}\}} 
F_\ab ~p_{(\al_1} \ldots p_{\al_{d-5})}
\Ga^I_\vs.
\label{specialx}
\eeq
Comparison of this expression 
with the form \rf{qf} of the operator $\Qhat_F$
confirms that $\de\Qhat$ and $\Qhat_F$
contain operators of the same general form.
So a choice of $\Xhat$ exists at each $d$ 
that creates from the conventional Dirac equation
a variety of linear spinor couplings to $F_\ab$.
Equivalently,
the corresponding terms in $\Qhat_F$
can be absorbed into other terms in the Lagrange density.

The other terms in $\cl_\ps$
that are associated with the redefinition \rf{specialx}
include ones involving different spinor couplings to $F_\ab$.
Notice that no operators involving 
only symmetrized covariant momenta can appear,
as $\Xhat$ is already linear in $F_\ab$.
Since $\Yhat = 0$ for the redefinition \rf{specialx}
and since the commutators \rf{pcomm}
imply that the result of $[p_\mu, \Xhat]$
is second order in $F_\ab$,
the terms involving different linear spinor couplings to $F_\ab$
arise from the anticommutator $p_\mu \{\ga^\mu, \Xhat\}$
with the extra momentum $p_\mu$ symmetrized 
with those in $\Xhat$.
This reveals that 
the resulting shift \rf{delQ} in $\Qhat$ 
contains terms at first order in $F_\ab$
given by
\bea
\de\Qhat &\supset&
i \m q 
\sum_{d>4}
X_I^{\vs\{\al\be\al_1 \ldots \al_{d-5}\}} F_\ab 
\Big( 
p_{(\al_1} \ldots p_{\al_{d-5})} \Ga^I_\vs
\nn\\
&&
\hskip 50pt
- \fr 1 {2\m} 
p_{(\mu} p_{\al_1} \ldots p_{\al_{d-5})} \{\ga^\mu, \Ga^I_\vs\}
\Big).
\nn\\
\label{frels}
\eea
Any given linear spinor coupling to $F_\ab$ at dimension $d$
is therefore paired with another at dimension $d+1$.
This implies that certain linear $F_\ab$ couplings
of mass dimension $d$ can be absorbed into others of dimension $d+1$.
The results are analogous to those obtained 
for the noninteracting case 
in Sec.\ II B of Ref.\ \cite{km13}. 

Note that other choices of $\Xhat$ can mix
linear spinor couplings to $F_\ab$ 
and operators with symmetrized covariant momenta,
via the commutator $[p_\mu, \Xhat] \ga^\mu$
and anticommutator $p_\mu \{\ga^\mu, \Xhat\}$ terms 
in Eq.\ \rf{delQ}.
This can be seen directly from Table \ref{fieldredefs}.
It implies more than one type of redefinition can be used 
to absorb certain linear spinor couplings to $F_\ab$,
which has potential consequences for the interpretation of models 
involving these couplings.

In the applications below to studies of Lorentz and CPT violation 
with Penning traps,
we keep all relevant terms rather than simplifying calculations
by absorbing some couplings via field redefinitions.
Although more labor intensive,
this reveals directly the combinations of measurable coefficients
and has the added benefit of permitting an extra check on calculations 
by verifying consistency with the redefinitions
shown in Table \ref{fieldredefs}.

\subsubsection{Scenario with constant $F_\ab$}
\label{Scenario}

For many experimental applications,
including those to Penning traps discussed
in the sections to follow,
the predominant part of the electromagnetic field strength
is constant in magnitude and direction in the laboratory frame.
In this scenario,
the Lagrange density $\cl_\ps$ presented 
in Sec.\ \ref{Lagrange density}
reduces to a simpler form for calculational purposes.
We remark in passing that a similar interpretation to what follows 
can also be envisaged for more general scenarios
involving nonconstant $F_\ab$,
whenever $F_\ab$ plays the role of a fixed background
rather than a dynamical field.

The requirement of constant $F_\ab$,
\beq
D_\ga F_\ab \equiv \prt_\ga F_\ab=0 ,
\eeq
immediately eliminates the contributions
$\cl^{(6)}_{\prt F}$ to $\cl_\ps$ presented in Eq.\ \rf{l6df}.
Moreover,
it also implies that the linear couplings to $F_\ab$
in the Lagrange densities \rf{l5f} and \rf{l6f}
can be reinterpreted in terms of simpler couplings
in the laboratory frame.
As an explicit example,
consider the coefficient $\acf 5 \mab$
appearing in $\cl^{(5)}_F$. 
This coefficient is contracted with $F_\ab$,
so when $F_\ab$ is constant the combination $\acf 5 \mab F_\ab$
effectively behaves like a contribution 
to the coefficient $\acm 3 \mu$ in the minimal Lagrange density \rf{lag3},
involving a coupling of mass dimension three instead of five.

This line of reasoning shows that 
most of the Lagrange densities $\cl^{(5)}_F$ and $\cl^{(6)}_F$ 
can be absorbed into the terms $\cl^{(3)}$ and $\cl^{(4)}$
when applied to scenarios with constant $F_\ab$,
via the replacements
\bea
\acm 3 \mu 
& \to &
\acm 3 \mu 
+ \half \acf 5 \mab F_\ab
,
\nn\\
\bcm 3 \mu 
& \to &
\bcm 3 \mu 
+ \half \bcf 5 \mab F_\ab
,
\nn\\
\Hcm 3 \mn 
& \to &
\Hcm 3 \mn 
+ \half \Hcf 5 \mnab F_\ab
,
\nn\\
\ccm 4 \ma
& \to &
\ccm 4 \ma 
+ \half \ccf 6 \mabc F_{\be\ga}
,
\nn\\
\dcm 4 \ma
& \to &
\dcm 4 \ma 
+ \half \dcf 6 \mabc F_{\be\ga}
,
\nn\\
\ecm 4 \al 
& \to &
\ecm 4 \al 
+ \half \ecf 6 \abc F_{\be\ga}
,
\nn\\
\fcm 4 \al 
& \to &
\fcm 4 \al 
+ \half \fcf 6 \abc F_{\be\ga}
,
\nn\\
\gcm 4 \mna 
& \to &
\gcm 4 \mna 
+ \half \gcf 6 \mnabc F_{\be\ga}.
\label{replacement}
\eea

The remaining terms involving the coefficients
$\mcf 5 \ab$ and $\mfcf 5 \ab$
can also be absorbed,
but into the fermion mass instead.
When $F_\ab$ is constant,
the combination $\mcf 5 \ab F_\ab$
represents a contribution to the Dirac mass,
while $\mfcf 5 \ab F_\ab$ 
acts as a chiral mass term.
The latter can be removed by a chiral transformation
with parameter $\th$ determined by
\beq
\ps \to e^{- i \th \ga_5} \ps,
\quad
\tan\th = \fr {\mfcf 5 \ab F_\ab} {2 \m + \mcf 5 \ab F_\ab}.
\eeq
This transformation leaves invariant the usual Dirac kinetic term,
and it has no leading-order effect on other Lorentz-violating terms 
because it differs from the identity
only by powers of coefficients for Lorentz violation.
The absorption of the coefficients $\mcf 5 \ab$ and $\mfcf 5 \ab$
is thereby found to be equivalent to the replacement
\bea
\m 
& \to & 
\sqrt{ (\m + \half \mcf 5 \ab F_\ab)^2 +(\half \mfcf 5 \ab F_\ab)^2} 
\nn\\
&\approx &
\m + \half \mcf 5 \ab F_\ab
\label{repl2}
\eea
at leading order in Lorentz violation.
Note that the coefficient $\mfcf 5 \ab$ 
is unobservable at this order.

The above discussion demonstrates that 
the spinor couplings to constant $F_\ab$ 
in Lorentz-violating QED with $d\leq 6$
reduce to terms in the minimal QED extension of Ref.\ \cite{ck}.
However,
the fermion mass and the minimal coefficients for Lorentz violation 
become dependent on $F_\ab$ 
according to the results \rf{replacement} and \rf{repl2}.
The operators with symmetrized covariant momenta
appearing in the Lagrange densities $\cl^{(5)}_D$ and $\cl^{(6)}_D$ 
are unaffected by this argument.

In the discussions below analyzing experiments with Penning traps,
the method described in this subsection
is used to simplify calculations with terms containing 
a factor of $F_\ab$.
As expected,
the results obtained are consistent with direct perturbative calculations
that explicitly keep the terms in $\cl_\ps$ 
involving spinor couplings to constant $F_\ab$.

\subsection{Frame changes}
\label{Frame changes}

This subsection outlines some generic considerations 
involving the frame changes that appear 
in performing an analysis for violations of rotation invariance.
More specific details for these and also other types of searches 
for Lorentz and CPT violation using Penning traps 
are provided in subsequent parts of this work. 

Tests of Lorentz and CPT symmetry with a trapped particle
effectively investigate its properties under rotations or boosts,
or compare its behavior to that of a trapped antiparticle. 
Since boosts close under commutation into rotations,
it is impossible to break Lorentz invariance 
without also breaking rotation invariance:
even if the physics predicted by a particular model 
is isotropic in a special frame,
any boost to another frame reintroduces anisotropic effects.
Also,
as CPT violation in realistic effective field theory
is accompanied by Lorentz violation
\cite{ck,owg},
it follows that CPT violation comes with rotation violation as well.
Tests of rotation symmetry are therefore 
of particular importance in the search for Lorentz and CPT violation.

The explicit form of a coefficient for Lorentz violation 
depends on the inertial frame of the observer.
Comparing different experiments thus 
involves comparing results in a standard frame.
The canonical frame adopted in the literature
is the Sun-centered celestial-equatorial frame 
\cite{sunframe},
which has the origin of its time coordinate $T$ 
defined as the 2000 vernal equinox.
The cartesian coordinates $X^J\equiv (X,Y,Z)$  in this frame
are specified as having the $Z$ axis 
aligned along the rotation axis of the Earth 
and the $X$ axis pointing from the Earth to the Sun,
with the $Y$ axis completing a right-handed coordinate system.
The Sun-centered frame is well suited as a standard frame
because it is essentially inertial during typical experimental time scales
and because its axes are conveniently chosen for laboratory studies. 

In any inertial frame in the vicinity of the Earth,
including the canonical Sun-centered frame,
the coefficients for Lorentz violation 
can be assumed to be constants in time and space
\cite{ck,akgrav}.
However,
the Earth rotates in this frame,
and so the coefficients for Lorentz violation
change with sidereal time when observed in the laboratory
\cite{ak98}.
As a result,
experimental observables for Lorentz violation
can oscillate in time at harmonics of the Earth's sidereal frequency
$\om_\oplus\simeq 2\pi/(23{\rm ~h} ~56{\rm ~min})$,
with their amplitudes and phases controlled by the coefficients.

To establish the time dependence of the coefficients
appearing in an experiment located on the Earth's surface,
it is useful to introduce a standard laboratory frame 
with time coordinate $t$ and 
cartesian coordinates $x^j \equiv (x,y,z)$
\cite{sunframe}.
The origin of $t$ can be defined conveniently for a given laboratory.
A useful choice is to match $t$ with the local sidereal time $T_\oplus$,
defined to have origin at a chosen moment 
when the $y$ axis lies along the $Y$ axis.
This is offset from the time $T$ in the Sun-centered frame by
any chosen integer number of sidereal rotations of the Earth
and by an additional shift 
\beq
T_0 \equiv T-T_\oplus 
\simeq 
\fr{(66.25^\circ- \la)}{360^\circ}
(23.934~{\rm hr}),
\label{T0}
\eeq
where $\la$ is the longitude of the laboratory in degrees.
The spatial axes in the standard laboratory frame are defined
with the $x$-axis pointing to local south, 
the $y$-axis pointing to local east,
and the $z$-axis pointing to the local zenith.
To obtain dominant effects,
both the boost $\be_\oplus\simeq 10^{-4}$ of the Earth
relative to the Sun-centered frame
and the boost $\be_L\simeq 10^{-6}$ of the laboratory
due to the rotation of the Earth
can be treated as negligible.
The relationship 
$x^{j} = R^{jJ} x^J$
between the coordinates $x^j$ in the laboratory frame
and the coordinates $x^J$ in the Sun-centered frame
is then given by the $T_\oplus$-dependent rotation matrix
\cite{sunframe}
\beq
R^{jJ}=\left(
\begin{array}{ccc}
\cos\ch\cos\om_\oplus T_\oplus
&
\cos\ch\sin\om_\oplus T_\oplus
&
-\sin\ch
\\
-\sin\om_\oplus T_\oplus
&
\cos\om_\oplus T_\oplus
&
0
\\
\sin\ch\cos\om_\oplus T_\oplus
&
\sin\ch\sin\om_\oplus T_\oplus
&
\cos\ch
\end{array}
\right).
\label{rotmat}
\eeq
This matrix generates the harmonic time dependences
of the coefficients for Lorentz violation 
observed in the laboratory frame.

For many laboratory experiments,
it is also convenient to introduce an apparatus frame
with cartesian coordinates 
$x^a \equiv (x^1 , x^2, x^3)$.
We denote the corresponding unit vectors by 
$(\hat x_1 , \hat x_2, \hat x_3)$.
For example,
in the experiments with Penning traps discussed below,
the $x^3$ axis is taken to be aligned with 
the uniform trapping magnetic field.
This may subtend a nonzero angle to the local zenith
specified in the standard laboratory frame by $\hat z$,
so that $\hat x_3 \cdot \hat z \neq 0$.
The relationship $x^a = R^{aj} x^j$
connecting the standard laboratory coordinates $(x,y,z)$ 
to the apparatus coordinates $(x^1,x^2,x^3)$
then involves a rotation matrix $R^{aj}$,
which can be specified in general
as the product of three Euler rotations
for suitable Euler angles $\al$, $\be$, and $\ga$,
\bea
R^{aj}(T) &=&
\left(
\begin{array}{ccc}
\cos\ga &\sin\ga  &0\\
-\sin\ga &\cos\ga &0\\
0 &0 &1\\
\end{array} 
\right)
\left(
\begin{array}{ccc}
\cos\be &0 &-\sin\be\\
0 &1 &0\\
\sin\be &0 &\cos\be\\
\end{array} 
\right)
\nn\\
&&
\times
\left(
\begin{array}{ccc}
\cos\al &\sin\al &0\\
-\sin\al &\cos\al &0\\
0 &0 &1\\
\end{array} 
\right).
\label{euler}
\eea
Combining the above results reveals that
the coordinates in the Sun-centered frame
are related to those in the apparatus frame by 
\beq
x^{a} (T_\oplus) 
= R^{a j} R^{jJ} (T_\oplus) X^J.
\label{transf}
\eeq
Expressions relating any given coefficients for Lorentz violation
in the two frames can be obtained from this result.

In addition to its sidereal rotation,
the Earth's revolution about the Sun induces further time variations
in the coefficients for Lorentz violation 
in the laboratory and apparatus frames.
These variations occur at harmonics of the annual frequency,
and they arise due to the boost $\be_\oplus \simeq 10^{-4}$
of the Earth in the Sun-centered frame.
Effects also arise from the boost $\be_L \simeq 10^{-6}$
of the laboratory due to the rotation of the Earth.
All these effects are suppressed by one or more powers of the boost.
Nonetheless,
they typically introduce experimental sensitivities
to coefficients for Lorentz violation 
beyond those observable via pure rotations,
as has been demonstrated in the literature
\cite{gkv14,kv15,ca04,he08,space}.
They can also be of larger magnitude than effects suppressed
by other mechanisms,
such as those involving couplings to electromagnetic fields in the apparatus. 
However,
to retain a reasonable scope for the present work,
we disregard boost effects in 
our analysis below of experiments with Penning traps,
focusing instead on sidereal signals arising from rotations. 
A treatment of boost effects for trapped particles
is feasible in principle 
and would make an excellent subject for a future work.

\section{Application to Penning traps}
\label{Application to Penning traps}

In this section,
the Lagrange density \rf{fermlag} is used 
as the starting point for an analysis
of the sensitivity to Lorentz and CPT violation  
attainable in experiments with Penning traps.
We apply perturbation theory 
to determine the shifts in energy levels 
for electrons, positrons, protons, and antiprotons
arising in the presence of coefficients for Lorentz violation.
This leads to expressions for the dominant shifts 
in the cyclotron and anomaly frequencies of trapped particles
and permits studies of experimental effects.
Observable signals arise from sidereal variations
and comparisons of particle and antiparticle properties.
We apply the results to published data from Penning traps
to obtain first constraints on several SME coefficients,
and we investigate the potential signals
in some forthcoming experiments.

\subsection{Theory}
\label{Theory for Penning trap}

For precision experiments on particles in Penning traps,
the transitions of primary interest
involve the energy levels created
by the constant magnetic field of the trap.
It is therefore appropriate to base the theoretical analysis 
on the idealized scenario of a relativistic charged quantum particle 
moving in a uniform magnetic field,
for which the unperturbed eigenenergies 
are the relativistic Landau levels.
The signals of experimental interest are energy-level shifts
rather than transition probabilities.
The dominant effects from Lorentz and CPT violation
can thus be treated as perturbative energy-level shifts,
to be added to the independent perturbations 
generated by radiative corrections in conventional QED
such as the splitting induced 
by the anomalous magnetic moment of the trapped particle.

The unperturbed eigenenergies and eigenfunctions
in the absence of Lorentz violation or radiative corrections 
can be found by solving the minimally coupled Dirac equation
for a spin-1/2 fermion
of mass $m$ and charge $q \equiv \Z |q|$ of fixed sign $\Z$
in a constant magnetic field.
For definiteness,
we choose the apparatus frame such that the magnetic field 
$\mbf B = B \hat x_3$ lies along the positive $x^3$ axis,
and we fix the gauge so that the electromagnetic potential 
is $A^\mu = (0,x_2 B,0,0) = (0,-x^2 B,0,0)$.
The spin-up and spin-down eigenstates form 
the two stacks of relativistic Landau levels,
which are degenerate except for the ground state.
We denote the level number as $n=0,1,2,3,\dots$
and label the fermion spin relative to the magnetic field
by $s=+1$ and $s=-1$ for up and down, respectively.
The stacks of levels are similar for the antifermion,
but with spin labels reversed.

At the $n$th level,
the stationary eigenstates $\ch_{n,s}$ for the positive-energy fermion 
are associated with eigenenergies
\beq
E_{n,s} = \sqrt{ m^2 + p_3^2 + (2n + 1 - \Z s) |qB|}.
\eeq 
The energy eigenvalues 
for the corresponding antifermion eigenstates $\ch_{n,s}^c$
take the same form,
but with $\Z$ being the opposite sign.
For example,
this equation encodes the information that 
the ground state for electrons or antiprotons is $E_{0,-1}$ with spin down,
while that for positrons or protons is $E_{0,+1}$ with spin up.

The four-component eigenspinors $\ch_{n,s}$
are given by
\bea
\ch_{n,+1}
&=& 
N_{n,+1}
\left( 
\begin{array}{c}
(m+E_{n,+1}) u_n
\\
0
\\ 
p_3 u_n 
\\ 
-\sqrt{|qB|} ~u_{n+1}
\end{array}
\right), 
\nn\\
\ch_{n,-1}
&=& 
N_{n,-1}
\left( 
\begin{array}{c}
0 
\\
(m+E_{n,-1}) u_n
\\
- 2n \sqrt{|qB|} ~u_{n-1}
\\
-p_3 u_n
\end{array}
\right) ,
\label{chi}
\eea
where the functions $u_n(\ze)$ are defined as
\beq
u_n (\ze) = 
\exp{(ip_1 x^1 +ip_3 x^3 )} \exp{(-\ze^2/2)} H_n(\ze),
\eeq
in terms of the Hermite polynomials $H_n(\ze)$,
with 
\beq
\ze = \sqrt{|qB|}\left( x^2 + \fr{p_1}{\Z|qB|} \right).
\eeq
The normalization factors $N_{n,s}$ are 
\beq
N_{n,s} =
\sqrt{ 
\fr{\sqrt{|qB|}}
{\sqrt{\pi} ~2^{n+1} n! ~E_{n,s} (m + E_{n,s}) L^2},
}
\eeq
where a cutoff $L$ has been adopted along the $x^1$ and $x^3$ directions.
The corresponding antifermion eigenstates 
$\ch^c_{n,s}$
are found to be
\bea
\ch^c_{n,+1}
&=& 
N_{n,+1}
\left( 
\begin{array}{c}
(m+E_{n,+1}) u_n
\\ 
0
\\
p_3 u_n 
\\ 
2 n \sqrt{|qB|} ~u_{n-1}
\end{array}
\right), 
\nn\\
\ch^c_{n,-1}
&=& 
N_{n,-1}
\left( 
\begin{array}{c}
0 
\\
(m+E_{n,-1}) u_n
\\
\sqrt{|qB|} ~u_{n+1}
\\
-p_3 u_n
\end{array}
\right) ,
\label{chic}
\eea
where the various quantities are defined as before
but involve the opposite value of $\Z$.
These positive-energy antifermion eigenstates can be obtained 
from the negative-energy fermion solutions
by charge conjugation in the usual way.

\subsubsection{Perturbative energy shift}
\label{Perturbative energy shift}

The unperturbed eigenstates \rf{chi} 
can be used to calculate the perturbative shifts
of the particle eigenenergies
once the perturbation hamiltonian $\de\cH$ is known.
However,
a direct construction of $\de\cH$
is challenging due to the higher powers of momenta
appearing in the Lorentz-violating operator $\Qhat$.
Following Ref.\ \cite{km12},
we can instead adopt a procedure
that yields an approximation to $\de\cH$ 
valid at leading order in Lorentz violation.

The exact hamiltonian $\cH$ can be defined
from the modified Dirac equation via 
\beq
(p^0- \cH) \ps = \ga_0 (p \cdot \ga - m + \Qhat) \ps = 0,
\eeq
where $p^0$ is the exact energy.
This gives
\beq
\cH = \ga_0 (\pvec \cdot \gavec + m - \Qhat) \equiv \cH_0+\de \cH,
\eeq
where 
$\de\cH = -\ga_0 \Qhat$ is the exact perturbation hamiltonian.
This form cannot be used directly in a perturbative calculation
because $\de\cH$ depends on the eigenenergies of $\cH$
and therefore requires prior knowledge of the energy shifts. 
However,
the energy shifts are perturbative,
so their contributions to $\de\cH$ 
lead to corrections at second or higher order 
in coefficients for Lorentz violation.
This means that leading-order results
can be derived by evaluating $\de\cH$
using the unperturbed eigenenergies,
\beq
\de\cH \approx -\ga_0 \Qhat |_{p^0\to E_{n,s}}.
\eeq
For a trapped fermion,
the dominant perturbative energy shifts 
due to Lorentz and CPT violation
are therefore given by the matrix elements
\beq
\de E_{n,s}= \vev{\ch_{n,s}|\de\cH|\ch_{n,s}}.
\label{deltae}
\eeq

The corresponding perturbation hamiltonian $\de\cH^c$
for antiparticles is obtained from $\de\cH$ 
by reversing the sign of the charge $q$ and the spin orientation $s$
and changing the sign of all coefficients for Lorentz violation
that control CPT-odd operators.
These coefficients are identified in Table \ref{nrcoeff}.
The shifts in the antiparticle energy levels
can then be obtained using the unperturbed eigenstates \rf{chic},
\beq
\de E^c_{n,s}= \vev{\ch^c_{n,s}|\de\cH^c |\ch^c_{n,s}}.
\eeq

To obtain explicit results,
we can take advantage of the constancy of the magnetic field
and adopt the approach presented in Sec.\ \ref{Scenario}.
It therefore suffices to limit attention 
to the operators appearing in 
$\cl^{(3)}$, $\cl^{(4)}$, $\cl^{(5)}_D$, and $\cl^{(6)}_D$.
After calculation with these terms,
we obtain the corresponding perturbative energy shifts.
The results are somewhat lengthy,
so here we report them only for $d=3$, 4, and 5
and relegate them to Appendix \ref{Energy shift}.
They hold at leading order in coefficients for Lorentz violation
but are exact in other quantities.
To obtain the additional contributions from operators in
$\cl^{(5)}_F$ and $\cl^{(6)}_F$,
it suffices to apply the substitutions
listed in Sec.\ \ref{Scenario},
while keeping only terms linear in coefficients for Lorentz violation.
The corresponding energy-level shifts for a trapped antiparticle
can be obtained from those for the particle
by reversing the spin $s$
and changing the signs $\si$ of the charge
and of all coefficients controlling CPT-odd operators.

The scales of all the energy shifts are set by the coefficients,
which are therefore the appropriate targets for experimental measurements.
However,
some contributions are suppressed.
Among these are corrections proportional to any nonzero power of $|qB|$,
all of which arise from operators involving covariant derivatives.
Even the comparatively large magnetic fields 
of $B \simeq 5$ T 
often found in Penning-trap experiments
produce only effects suppressed 
by $|eB|/m_e^2 \simeq 10^{-9}$ for electrons or positrons
and by $|eB|/m_p^2 \simeq 10^{-16}$ for protons or antiprotons.
This means that the results presented in Appendix \ref{Energy shift}
could be used together with experimental data
to obtain constraints on many coefficients associated with a factor of $|qB|$,
albeit yielding weaker sensitivities. 
However, 
since these coefficients are associated 
with covariant-derivative couplings,
they are also accessible in unsuppressed experimental studies 
of the behavior of free particles.
We therefore disregard effects proportional to $|qB|$ 
in what follows.
In contrast,
terms proportional to $B$ without a factor of $q$,
which arise from  
operators in $\cl^{(5)}_F$ and $\cl^{(6)}_F$,
represent Lorentz-violating couplings
that are independent of free-particle motion
and hence can only be detected in the presence of an electromagnetic field.
Comparatively few investigations of these terms
have been performed to date.
We therefore include these effects in this work,
placing first constraints on some of the coefficients
appearing in $\cl^{(5)}_F$ and $\cl^{(6)}_F$.

A Penning trap includes not only 
the radially confining magnetic field of uniform magnitude $B$ 
but also an axially confining electric field of varying magnitude $E$.
The Landau momentum $p_3$ appearing
in the expressions in Appendix \ref{Energy shift}
therefore physically represents an effective momentum
for the axial motion.
In the presence of the electric field,
terms involving powers of $p_3$
become expectation values of the physical axial momentum.
For a trapped particle the odd powers must vanish,
but the even powers can be expected to contribute.
When the axial quantum number is low,
neglecting energy shifts from the even powers is a reasonable approximation
because the ratio of axial to cyclotron frequencies
is typically much less than one.
Some cooling procedures may equipartition the axial and cyclotron energies
and thus lead to large axial quantum numbers,
which could produce Lorentz-violating perturbative shifts
proportional to $|qE|$
comparable to those proportional to $|qB|$. 
For coefficients associated with covariant-derivative couplings,
neglecting both effects is therefore consistent.
For the $F$-type coefficients involving $E$,
the effects are interesting in principle
because they cannot be studied in the absence of the electric field.
However,
they are more challenging to analyze because $E$ varies with position,
and moreover the sensitivity to these coefficients is typically weaker 
by the ratio $|E/B|$.
For example,
for a typical configuration with 100~V applied over about 5~mm
in a 5~T magnetic field,
the ratio is about $10^{-5}$ in natural units.
We therefore choose to disregard effects 
involving these coefficients in what follows.

With the above choices,
the perturbative energy shift $\de E_{n,s}$
for a fermion of species $w$,
charge sign $\si$, and spin sign $s$ in a magnetic field 
of magnitude $B$ oriented along $\hat x_3$ 
is found to have the form
\beq
\de \ens^w =
\atw w 0
-\Z s \btw w 3
-\mftw w 3 B
+ \Z s \bftw w {33} B  
\label{deleb}
\eeq
in the noninertial laboratory frame,
where the tilde quantities are convenient combinations 
of the cartesian coefficients, 
defined by 
\bea
\atw w 0
&=&
\acmw 3 0 w
- \mw \ccmw 4 {00} w
- \mw \ecmw 4 0 w
+ \mw^2 \mcw 5 {00} w
+ \mw^2 \acw 5 {000} w
\nn\\
&&
- \mw^3 \ccw 6 {0000} w
- \mw^3 \ecw 6 {000} w,
\nn\\
\btw w 3
&=&
\bcmw 3 3 w
+ \Hcmw 3 {12} w
- \mw \dcmw 4 {30} w
- \mw \gcmw 4 {120} w
\nn\\
&&
+ \mw^2 \bcw 5 {300} w
+ \mw^2 \Hcw 5 {1200} w
\nn\\
&&
- \mw^3 \dcw 6 {3000} w
- \mw^3 \gcw 6 {12000} w,
\nn\\
\mftw w 3
&=&
\mcfw 5 {12} w
+ \acfw 5 {012} w
- \mw \ccfw 6 {0012} w
- \mw \ecfw 6 {012} w,
\nn\\
\bftw w {33}
&=&
\bcfw 5 {312} w
+ \Hcfw 5 {1212} w
- \mw \dcfw 6 {3012} w
- \mw \gcfw 6 {12012} w .
\nn\\
\eea
The fermion-flavor dependence of the coefficients 
is reflected in the subscript $w$,
which can take the values $e$, $p$, 
and in principle others as well.
The indices 0, 3, and 33 on these tilde quantities 
correctly reflect their properties under spatial rotations,
as the index pair 12 is antisymmetric
wherever it appears on the right-hand side
and hence transforms like a single 3 index.
The dependence on only the $\hat x_3$ direction
is due to the cylindrical symmetry of the Penning trap.

We remark in passing that the first four terms of the quantity $\btw w 3$ 
form a widely used coefficient in studies of the minimal SME,
also denoted $\btw w 3$
\cite{tables},
which here is extended to include $d=5$ and 6 effects.
In fact,
judicious use of Eqs.\ (26)-(28) of Ref.\ \cite{km13}
permits a further generalization of this coefficient
to include effects arising at arbitrary $d$,
giving
\bea
\btw w 3 &=&
\sum_{d}
\mw^{d-3} 
(
\bcw d {30^{d-3}} w 
+ \Hcw d {120^{d-3}} w
\nn\\[-10pt]
&&
\hskip 40pt
- \dcw d {30^{d-3}} w
- \gcw d {120^{d-3}} w
),
\label{allorders}
\eea
where the sum is over odd values of $d$ for 
the $b$- and $H$-type coefficients
and over even values of $d$ for 
the $d$- and $g$-type coefficients, 
and where the index $0^{d-3}$ 
denotes $d-3$ timelike indices.
A similar result can be obtained for $\atw w 0$.
Obtaining the analogous expressions for the $F$-type coefficients
requires the Lagrange density for $F$-type couplings at arbitrary $d$,
which remains unexplored to date.

The perturbative energy shift \rf{deleb}
is the key to extracting dominant signals for Lorentz and CPT violation
in Penning-trap experiments.
It reveals that only four quantities in the noninertial laboratory frame,
the tilde coefficients 
$\atw w 0$, $\btw w 3$, $\mftw w 3$, and $\bftw w {33}$,
govern all the dominant Lorentz-violating energy shifts
for a given fermion in an idealized Penning trap.
However, 
the isotropic coefficient $\atw w 0$ provides 
the same instantaneous shift for all energy levels,
which cancels in all frequencies and is therefore unobservable.
Moreover,
the coefficient $\mftw w 3$ also provides an identical instantaneous shift
to all energy levels,
despite the shift being dependent on the magnetic field
and ultimately also dependent on sidereal time 
due to the coefficient anisotropy.
In contrast,
the other two coefficients $\btw w 3$ and $\bftw w {33}$
can in principle be detected in suitable experiments.
They contribute with opposite signs for spin up and spin down
and therefore shift the two Landau-level stacks
relative to each other,
which is a measurable effect.
This shift preserves the level spacing within each stack
because the perturbation \rf{deleb} is independent 
of the level number $n$.
Notice that the magnitude $|q|$ of the fermion charge
plays no role here.

The expression for the perturbative energy shift $\de \ens^{\ol w}$
for the corresponding antifermion 
is obtained by reversing the orientation of the spin $s$
and the signs of the coefficients controlling CPT-odd operators
in the energy shift \rf{deleb},
\bea
\de \ens^{\ol w} 
&=& 
\de \enms
|_{(a,b,e,g)\to (-a,-b,-e,-g)}
\nn\\
&\equiv& 
-\atws w 0
+\Z s \btws w 3
-\mftws w 3 B
- \Z s \bftws w {33} B ,
\label{delebbar}
\eea
where the set of four starred tilde coefficients 
is defined by
\bea
\atws w 0
&=&
\acmw 3 0 w
+ \mw \ccmw 4 {00} w
- \mw \ecmw 4 0 w
- \mw^2 \mcw 5 {00} w
+ \mw^2 \acw 5 {000} w
\nn\\
&&
+ \mw^3 \ccw 6 {0000} w
- \mw^3 \ecw 6 {000} w,
\nn\\
\btws w 3
&=&
\bcmw 3 3 w
- \Hcmw 3 {12} w
+ \mw \dcmw 4 {30} w
- \mw \gcmw 4 {120} w
\nn\\
&&
+ \mw^2 \bcw 5 {300} w
- \mw^2 \Hcw 5 {1200} w
\nn\\
&&
+ \mw^3 \dcw 6 {3000} w
- \mw^3 \gcw 6 {12000} w,
\nn\\
\mftws w 3
&=&
\mcfw 5 {12} w
- \acfw 5 {012} w
- \mw \ccfw 6 {0012} w
+ \mw \ecfw 6 {012} w,
\nn\\
\bftws w {33}
&=&
\bcfw 5 {312} w
- \Hcfw 5 {1212} w
+ \mw \dcfw 6 {3012} w
- \mw \gcfw 6 {12012} w .
\nn\\
\eea
In deriving the result \rf{delebbar},
the sign $\si$ of the fermion charge
is understood to change,
the orientation of the magnetic field is assumed constant,
and the direction $s$ of the spin is still taken
relative to the magnetic field.
In parallel with the fermion case,
only the combinations $\btws w 3$ and $\bftws w {33}$
are observable.

\subsubsection{Cyclotron and anomaly frequencies}
\label{Cyclotron and anomaly frequencies}

In Penning-trap experiments,
the primary observables are frequencies.
Two key frequencies are
the cyclotron frequency $\nu_c = \om_c/2\pi$
and the Larmor spin-precession frequency $\nu_L = \nu_a + \nu_c$,
where $\nu_a = \om_a/2\pi$ is the anomaly frequency
\cite{geo}.
In the presence of Lorentz and CPT violation,
these frequencies can become shifted.
For experiments with a fixed magnetic field
and trapped fermions or antifermions of a given flavor $w$,
the dominant shifts depend on only the four combinations
$\btw w 3$, $\bftw w {33}$, $\btws w 3$, and $\bftws w {33}$
of cartesian coefficients in the noninertial laboratory frame.
In this subsection,
we use the results \rf{deleb} and \rf{delebbar}
to determine these shifts 
for the cyclotron and anomaly frequencies
of trapped electrons, positrons, protons, and antiprotons.
We show that the shifts are governed 
by a total of 36 independent inertial-frame observables
in Penning-trap experiments,
formed as combinations of 432 independent components 
of cartesian coefficients in the Sun-centered frame.

The cyclotron frequency $\om_c$ is in natural units
the energy difference between the ground-state $n=0$ level
and the $n=1$ level in the same Landau stack,
which for the particles of interest here is the stack with $s=\si$.
Since the perturbations \rf{deleb} and \rf{delebbar}
are are independent of $n$ 
and therefore constant for fixed $s$ and $\si$,
no change in the cyclotron frequency appears at leading order,
\bea
\de \om_c^{w} &=&
\de E_{1,\Z}^{w}-\de E_{0,\Z}^{w} 
\approx 0,
\nn\\
\de \om_c^{\ol w} &=&
\de E_{1,\Z}^{\ol w}-\de E_{0,\Z}^{\ol w} 
\approx 0 ,
\label{cyclnochange}
\eea
for either a fermion $w = e^-$, $p$
or for an antifermion $\ol w = e^+$, $\ol p$.
Note that the exact expressions for the energy shifts
in Appendix \ref{Energy shift}
reveal the existence of subleading effects suppressed by $|qB|$
that do vary with $n$ and therefore 
can produce subleading shifts in the cyclotron frequency,
but these can be neglected here in accordance with 
the discussion in the previous subsection.

The dominant Lorentz-violating effects thus appear
as shifts in the anomaly frequency $\om_a$.
In natural units and for the particles relevant here,
this is the energy difference
between the $n=1$ level in the Landau stack with $s=\si$
and the $n=0$ level in the stack with $s=-\si$.
Using the perturbative corrections \rf{deleb} and \rf{delebbar}
reveals that the anomaly frequencies 
for either a fermion $w = e^-$, $p$
or for an antifermion $\ol w = e^+$, $\ol p$
are shifted according to 
\bea
\de \om_a^{w} 
&=&
\de E_{0,-\si}^{w}-\de E_{1,\si}^{w}
= 2 \btw w 3 - 2 \bftw w {33} B ,
\nn\\
\de \om_a^{\ol w}
&=&
\de E_{0,-\si}^{\ol w}-\de E_{1,\si}^{\ol w} 
= - 2 \btws w 3 + 2 \bftws w {33} B .
\label{delomaw}
\eea
Note that for each flavor
all four tilde coefficients in the laboratory frame 
appear in these expressions.
Note also that the antifermion result
can be obtained from the fermion one
by changing the signs of all the basic coefficients
associated with CPT-odd operators,
as might be expected.

The above formulae for the shifts in the anomaly frequencies
involve coefficients controlling a mixture of
CPT-even and CPT-odd effects.
However,
comparisons between particles and antiparticles
can in principle
permit the independent extraction of the CPT-odd contributions.
For simplicity,
suppose the magnetic fields in the two measurements
have the same magnitude and orientation.
Given the shifts in the anomaly frequencies
$\de \om_a^{w}$ for a fermion 
and $\de \om_a^{\ol w}$
for its antifermion,
we can take the difference to obtain
\bea
\De \om_a^{w} &\equiv&
\half (\de \om_a^{w} - \de \om_a^{\ol w})
\nn\\
&=& \btw w 3 - \bftw w {33} B 
+ \btws w 3 - \bftws w {33} B
\nn\\
&=&
2 \bcmw 3 3 w
- 2 \mw \gcmw 4 {120} w
+ 2 \mw^2 \bcw 5 {300} w
- 2 \mw^3 \gcw 6 {12000} w
\nn\\
&&
- 2 \bcfw 5 {312} w B
+ 2 \mw \gcfw 6 {12012} w B
\nn\\
&=&
2\De\btw w 3 + 2\De\bftw w {33},
\label{Deltaom}
\eea
where in the last expression
we have introduced the convenient definitions 
\bea
\De\btw w 3 &\equiv& 
\half ( \btw w 3 - \btws w 3 )
\nn\\
&=&
\bcmw 3 3 w
- \mw \gcmw 4 {120} w
+ \mw^2 \bcw 5 {300} w
- \mw^3 \gcw 6 {12000} w,
\nn\\
\De\bftw w {33} &\equiv& 
\half ( \bftw w {33} - \bftws w {33} )
\nn\\
&=&
- \bcfw 5 {312} w 
+ \mw \gcfw 6 {12012} w .
\label{debdefs}
\eea
The result \rf{Deltaom} shows explicitly 
that only coefficients for CPT violation 
appear in $\De \om_a^{w}$.
In fact,
all of the CPT-odd effects are encoded in the difference $\De \om_a^{w}$,
as the orthogonal combination
\bea
\Si \om_a^{w} &\equiv&
\half (\de \om_a^{w} + \de \om_a^{\ol w})
\nn\\
&=& 
\btw w 3 - \bftw w {33} B 
- \btws w 3 + \bftws w {33} B 
\nn\\
&=&
2  \Hcmw 3 {12} w
- 2 \mw \dcmw 4 {30} w
+ 2 \mw^2 \Hcw 5 {1200} w
- 2 \mw^3 \dcw 6 {3000} w
\nn\\
&&
+ 2 \Hcfw 5 {1212} w B
- 2 \mw \dcfw 6 {3012} w B
\label{Sumom}
\eea
contains only coefficients for CPT-even Lorentz violation. 
We remark in passing that each term contributing to the CPT violation 
in the result \rf{Deltaom} is also CT violating,
as predicted by the discussion in Sec.\ II C of Ref.\ \cite{bkr98}.

To express the shifts \rf{delomaw} in the anomaly frequencies
and the difference \rf{Deltaom}
in terms of constant coefficients in the Sun-centered frame
requires applying the methods described in Sec.\ \ref{Frame changes}.
This thereby reveals the sidereal-time and geometric dependences
of the laboratory-frame tilde coefficients.
As a simple example,
consider a scenario having the laboratory located at colatitude $\ch$
with the magnetic field pointing to the local zenith
so that $\hat x_3 = \hat z$,
and focus on the single-index laboratory-frame coefficient $\btw w 3$.
In this special case,
application of the rotation matrices given 
in Sec.\ \ref{Frame changes}
and the transformation \rf{transf}
yields the result
\beq
\btw w 3 =
\btw w Z \cos\ch 
+ ( \btw w X \cos\om_\oplus T_\oplus 
+ \btw w Y \sin\om_\oplus T_\oplus )\sin\ch ,
\eeq
expressing the noninertial-frame quantity $\btw w 3$
in terms of the three independent quantities $\btw w J$, $J=X,Y,Z$, 
in the canonical inertial frame.
More generally,
when the magnetic field points along a generic direction
in the laboratory frame,
trigonometric functions of the extra Euler angles
$\al$, $\be$, $\ga$ in Eq.\ \rf{euler}
appear as well.
 
In an analogous fashion,
the laboratory-frame tilde coefficient 
$\bftw w {33}$
is associated with the six independent 
combinations $\bftw w {(JK)}$ in the Sun-centered frame.
These produce up to second harmonics
in the sidereal frequency,
due to the nature of $\bftw w {(JK)}$ as an observer 2-tensor.
For example,
in the above simple scenario at colatitude $\ch$
with the magnetic field pointing to the local zenith,
we find
\bea
\bftw w {33}
&=&
\bftw w {ZZ}
+\half (\bftw w {XX} +\bftw w {YY} -2\bftw w {ZZ}) \sin^2\ch
\nn\\ 
&&
+( \bftw w {(XZ)} \cos\om_\oplus T_\oplus
+ \bftw w {(YZ)} \sin\om_\oplus T_\oplus )\sin 2\chi 
\nn\\ 
&&
+[
\half (\bftw w {XX} - \bftw w {YY}) \cos 2\om_\oplus T_\oplus
\nn\\ 
&&
\hskip 45pt
+ \bftw w {(XY)} \sin 2\om_\oplus T_\oplus ] \sin^2\ch .
\eea
Taking into account the relevant two fermion flavors $w$
and including also experiments with antiparticles,
which can access the nine additional independent combinations
$\btws w J$ and $\bftws w {(JK)}$,
we can conclude that there are 36 independent tilde observables
in the Sun-centered frame.
Each of these observables is 
a linear combination of cartesian coefficients,
of which 12 independent components appear 
in the perturbative corrections \rf{deleb} and \rf{delebbar}.
Various combinations such as the 18 independent differences
\bea
\De\btw w J &\equiv& \half ( \btw w J - \btws w J ),
\nn\\
\De\bftw w {(JK)} &\equiv& 
\half ( \bftw w {(JK)} - \bftws w {(JK)} )
\eea
may also appear in performing experimental analyses.
We thus see that the 36 independent observables
in Penning-trap experiments
are formed as linear combinations of 432 independent components
of cartesian coefficients in the Sun-centered frame.
Each observable corresponds to a physically distinct
and dominant Lorentz-violating effect,
so Penning traps offer excellent coverage
of the available coefficient space,
and moreover coverage at high sensitivity. 

Any single Penning-trap experiment 
with fixed magnetic field and a given particle 
can in principle access four harmonics and a constant term,
although the latter is time independent and hence challenging to measure.
This means that at most five of the 36 independent pieces of information
are accessible in any given experiment.
A joint analysis of data from multiple experiments 
is therefore required to explore fully the available coefficient space.
Complete coverage can be obtained 
only if experiments are performed
with all relevant particle and antiparticles
and if different experimental geometries are adopted.
The experimental conditions can be changed by changing the orientation
or magnitude of the magnetic field,
or by performing the experiment at a different colatitude.

\subsection{Experiments}
\label{Experiments}

In this subsection,
we first discuss some concepts
essential to studies of Lorentz and CPT violation
in Penning-trap experiments.
These concepts and the results obtained above
are then used to extract estimated constraints on
coefficients for Lorentz and CPT violation
and to predict potential future signals
in some existing and forthcoming experiments.

\subsubsection{Concepts}
\label{Concepts}

Studies of the anomalous magnetic moment and the $g$ factor
of a particle in a Penning trap 
can be idealized as measurements of the ratio 
of the anomaly frequency $\nu_a$
to the cyclotron frequency $\nu_c$,
linked to $g$ in a Lorentz- and CPT-invariant scenario by 
\beq
\fr {\nu_a}{\nu_c} \equiv \fr {\om_a}{\om_c} = \fr {g}{2} -1
\quad 
{\rm (Lorentz/CPT~invariance)}.
\label{ratio}
\eeq
In this conventional Lorentz- and CPT-invariant case,
$g$ is a numerical scalar quantity 
that is an intrinsic property of the particle.
The predicted value of $g$ can in principle be calculated
in a suitable theoretical framework
such as Lorentz-invariant quantum field theory,
and it is related to fundamental quantities
such as the fine structure constant. 
Radiative corrections modify the theoretical tree-level value of $g$
\cite{eradcorr},
and real measurements must take into account
various experimental effects involving the axial frequency, 
the relativistic shift, the cavity shift, and more
\cite{exptshift},
but $g$ remains an intrinsic numerical property of the particle.

In the presence of Lorentz and CPT violation,
this scenario is drastically changed
because the energies and hence the anomaly frequency $\om_a$
are directly shifted,
as is evident from Eq.\ \rf{deleb}.
The portion of the shift associated with the coefficient $\btw w 3$
is independent of the magnitude of $B$
but depends on geometric factors 
such as the local sidereal time $T_\oplus$,
the colatitude $\ch$ of the experiment,
and the direction $\hat x_3$ of the magnetic field,
while the part involving $\bftw w {33}$
depends both on $B$ and on the geometric factors.
In short,
the anomaly frequency can be viewed as a function
of these variables,
\beq
\om_a = \om_a (T_\oplus,\ch,\hat x_3,B)
\quad 
{\rm (Lorentz/CPT~violation)}.
\label{omalv}
\eeq
An immediate consequence is that
{\it the experimental ratio $\om_a/\om_c$ 
is no longer an intrinsic property of the particle}
and instead becomes an experiment-dependent quantity.
Reported values of $g$ obtained using the result \rf{ratio} 
therefore cannot be directly compared between experiments
in a meaningful way
because they depend on the local experimental conditions:
the local sidereal time,
the colatitude of the laboratory,
and the direction and magnitude of the magnetic field.
Instead,
the intrinsic quantities that provide experiment-independent
measures of Lorentz and CPT violation
are SME coefficients expressed in the canonical Sun-centered frame.
In the present context of Penning-trap experiments
with electrons, positrons, protons, and antiprotons,
these intrinsic quantities can be taken as the 36 tilde coefficients
$\btw e J$, $\btws e J$, $\bftw e {(JK)}$, $\bftws e {(JK)}$
and 
$\btw p J$, $\btws p J$, $\bftw p {(JK)}$, $\bftws p {(JK)}$
in the Sun-centered frame.
They can be extracted from the ratios $\om_a/\om_c$
obtained for the different species
under various laboratory conditions,
by matching to the predicted dependences
on the geometrical factors relevant for each given experiment. 

In the conventional context with Lorentz and CPT invariance,
one experimental advantage of extracting the ratio \rf{ratio}
is that both $\om_a$ and $\om_c$ are proportional to $B$,
so $B$ cancels in the determination of $g$.
For example,
if the measurements can be performed quasi-simultaneously,
then accurate knowledge of $B$ is unnecessary 
to achieve a high-precision measurement of $g$.
However,
in the presence of Lorentz and CPT violation,
the ratio \rf{ratio} is no longer independent of $B$
because the coefficients $\btw w J$, $\btws w J$
appear without an accompanying factor of $B$.
A precision measurement of these coefficients 
therefore requires continous calibration of $B$
as implemented,
for instance,  
in a sidereal-variation analysis 
performed at the University of Washington
\cite{mi99}.
However,
for measurements restricting attention to the $F$-type coefficients
$\bftw w {(JK)}$ and $\bftws w {(JK)}$,
which always come with a factor of $B$,
the cancellation remains in force
and accurate knowledge of $B$ is again unnecessary.

\begin{table*}
\caption{
\label{geometry}
Geometrical quantities for some experiments.}
\setlength{\tabcolsep}{7pt}
\begin{tabular}{ccccccc}
\hline
\hline
Experiment	&	Species	&		$\ch$		&	$\hat x_3$	&	$B$	&	$\la$	&	$T_0$	\\	\hline
Washington \cite{de99}	&	$e^-$, $e^+$	&	$	42.5^{\circ}	$	&	upward	&	5.85 T	&	$-122.3^\circ$	&	12.54 h	\\	
Washington \cite{mi99}	&	$e^-$	&	$	42.5^{\circ}	$	&	upward	&	5.85 T	&	$-122.3^\circ$	&	12.54 h	\\	
Harvard \cite{ha11}	&	$e^-$	&	$	47.6^{\circ}	$	&	upward	&	5.36 T	&	$-71.1^\circ$	&	9.13 h	\\	
Harvard \cite{fo15}	&	$e^+$	&	$	47.6^{\circ}	$	&	upward	&	$\simeq 6$ T	&	$-71.1^\circ$	&	9.13 h	\\	
ATRAP \cite{di13}	&	$\ol p$	&	$	43.8^{\circ}	$	&	upward	&	5.2 T	&	$6.1^\circ$	&	4.00 h	\\	
BASE \cite{mo14}	&	$p$	&	$	40.0^{\circ}	$	&	south	&	1.90 T	&	$8.3^\circ$	&	3.85 h	\\	
BASE \cite{chip}	&	$\ol p$	&	$	43.8^{\circ}	$	&	$60^{\circ}$ west of north	&	1.95 T	&	$6.1^\circ$	&	4.00 h	\\	
\hline
\hline
\end{tabular}
\end{table*}

Implications related to the above conceptual points
also arise for comparative tests
involving particles and antiparticles.
Suppose one experiment measures the ratio
${\om_a^w}/{\om_c^w}$ for a particle of species $w$,
while a second experiment measures the ratio
${\om_a^{\ol w}}/{\om_c^{\ol w}}$ 
for the corresponding antiparticle.
We are allowing here for the possibility that the cyclotron frequencies 
$\om_c^w$, $\om_c^{\ol w}$ 
of the two measurements may differ 
due to different magnitudes of the experimental magnetic fields.
In a Lorentz- and CPT-invariant scenario,
the difference between these two measurements is 
\beq
\fr {\om_a^w}{\om_c^w} - \fr {\om_a^{\ol w}}{\om_c^{\ol w}} 
=\half (g - \ol g) 
\quad 
{\rm (CPT~invariance)}
\label{omdifference}
\eeq
according to Eq.\ \rf{ratio}.
The CPT theorem guarantees that this quantity is identically zero.

However,
in the presence of CPT violation,
the picture again changes drastically
due to the qualitatively different nature 
of the anomaly frequency \rf{omalv}.
Using Eq.\ \rf{delomaw},
the difference between the two measurements 
is found to be
\beq
\fr {\om_a^w}{\om_c^w} - \fr {\om_a^{\ol w}}{\om_c^{\ol w}} =
\fr {\de\om_a^w}{\om_c^w} - \fr {\de\om_a^{\ol w}} {\om_c^{\ol w}} 
\quad 
{\rm (CPT~violation)},
\label{omdiff}
\eeq
since the CPT theorem guarantees the cancellation
of all Lorentz- and CPT-invariant contributions.
We see from the result \rf{omalv} that 
{\it the experimental difference
$(\om_a^w/\om_c^w)-(\om_a^{\ol w}/\om_c^{\ol w})$
depends on the local experimental conditions:}
the local sidereal time,
the colatitudes of the laboratories 
where the two experiments are performed,
and the directions and magnitudes of the magnetic fields.

More insight can be gained by algebraically expressing
the difference \rf{omdiff}
in terms of sums and differences
of the anomaly and cyclotron frequencies,
defined as
\bea
\De \om_a^{w} &=&
\half (\de \om_a^{w} - \de \om_a^{\ol w}),
\quad
\Si \om_a^{w} =
\half (\de \om_a^{w} + \de \om_a^{\ol w}),
\nn\\
\De \om_c^{w} &=&
\half (\om_c^{w} - \om_c^{\ol w}),
\quad
\Si \om_c^{w} =
\half (\om_c^{w} + \om_c^{\ol w}).
\eea
This gives
\bea
\fr {\om_a^w}{\om_c^w} - \fr {\om_a^{\ol w}}{\om_c^{\ol w}} 
&=&
\fr 2 {\om_c^w \om_c^{\ol w}} 
\left( \Si \om_c^{w} \De \om_a^{w} 
- \De \om_c^{w} \Si \om_a^{w}\right).
\label{sumdiff}
\eea
We have seen in the previous section that
no leading-order changes in the cyclotron frequencies occur 
in the presence of Lorentz and CPT violation,
so any difference $\De \om_c^{w}$ is purely due to
experimental magnetic fields of different magnitude.
For magnetic fields of identical orientation 
the theoretical predictions
\rf{Deltaom} for $\De \om_a^{w}$
and \rf{Sumom} for $\Si \om_a^{w}$
show that the first term of the result \rf{sumdiff}
involves CPT violation,
while the second involves CPT-invariant Lorentz violation.
These points reveal that 
{\it the experimental difference
$(\om_a^w/\om_c^w)-(\om_a^{\ol w}/\om_c^{\ol w})$
is a clean measure of CPT violation
only if both measurements use magnetic fields
of identical strength and orientation.}

In the event that indeed both ratio measurements are
made using the same $\mbf B$, 
which implies $\om_c^w = \om_c^{\ol w}$,
then the explicit form of the difference \rf{sumdiff}
reduces to 
\bea
\fr {\om_a^w}{\om_c^w} - \fr {\om_a^{\ol w}}{\om_c^{\ol w}} 
&=&
\fr {2\De\om_a^w}{\om_c^w} 
\quad  
({\rm CPT~violation,} ~\om_c^{\ol w} = \om_c^w)
\nn\\
&=&
\fr {4} {\om_c^w} (\De\btw w 3 + \De\bftw w {33} B)
\label{equalomdiff}
\eea
by using the result \rf{Deltaom}.
This is is indeed a pure CPT test,
as only coefficients for CPT violation
enter the definitions \rf{debdefs}
for $\De\btw w 3$ and $\De\bftw w {33}$.
Conversion of this expression from the noninertial laboratory frame
to the canonical inertial Sun-centered frame
using the transformation \rf{transf}
displays the dependence on 
the 18 intrinsic experiment-independent observables
$\De\btw w J$, $\De\bftw w {(JK)}$
for CPT violation
and exposes the explicit dependence
on the local sidereal time $T_\oplus$,
the colatitude $\ch$ of the laboratory,
and the direction $\hat z$ of the magnetic field.

\renewcommand{\arraystretch}{1.5}
\begin{table*}
\caption{
\label{electronsector}
Analysis for the electron sector.}
\setlength{\tabcolsep}{7pt}
\begin{tabular}{cccc}
\hline
\hline
Experiment	&		Lab.\ frame		&			Sun-centered frame				&		Harmonic		\\	\hline
Washington \cite{de99}	&	$	\De\btw e 3	$	&	$	0.7	\De\btw e Z			$	&	$	1	$	\\	
	&	$	\De\bftw e {33}	$	&	$	0.2	(\De\bftw e {XX} + \De\bftw e {YY}) +	0.5	 \De\bftw e {ZZ}	$	&	$	1	$	\\	[5pt]
Washington \cite{mi99}	&	$	\btw e 3	$	&	$	0.7	\btw e X			$	&	$	\codt	$	\\	
	&	$		$	&	$	0.7	\btw e Y			$	&	$	\sodt	$	\\	
	&	$	\bftw e {33}	$	&	$		\bftw e {(XZ)} 			$	&	$	\codt	$	\\	
	&	$	  	$	&	$		\bftw e {(YZ)} 			$	&	$	\sodt	$	\\	[5pt]
Harvard \cite{ha11}	&	$	\btw e 3	$	&	$	0.7	\btw e X			$	&	$	\codt	$	\\	
	&	$		$	&	$	0.7	\btw e Y			$	&	$	\sodt	$	\\	
	&	$	\bftw e {33}	$	&	$		\bftw e {(XZ)} 			$	&	$	\codt	$	\\	
	&	$	  	$	&	$		\bftw e {(YZ)} 			$	&	$	\sodt	$	\\	
	&	$		$	&	$	0.3	(\bftw e {XX} - \bftw e {YY} )			$	&	$	\ctodt	$	\\	
	&	$		$	&	$	0.5	\bftw e {(XY)} 			$	&	$	\stodt	$	\\	[5pt]
Harvard \cite{fo15}	&	$	\btws e 3	$	&	$	-0.7	\btws e X			$	&	$	\codt	$	\\	
	&	$		$	&	$	-0.7	\btws e Y			$	&	$	\sodt	$	\\	
	&	$	\bftws e {33}	$	&	$	-	\bftws e {(XZ)} 			$	&	$	\codt	$	\\	
	&	$	  	$	&	$	-	\bftws e {(YZ)} 			$	&	$	\sodt	$	\\	
	&	$		$	&	$	-0.3	(\bftws e {XX} - \bftws e {YY} )			$	&	$	\ctodt	$	\\	
	&	$		$	&	$	-0.5	\bftws e {(XY)} 			$	&	$	\stodt	$	\\	
	&	$	\De\btw e 3	$	&	$	0.7	\De\btw e Z			$	&	$	1	$	\\	
	&	$	\De\bftw e {33}	$	&	$	0.3	(\De\bftw e {XX} + \De\bftw e {YY})  +	0.5	\De\bftw e {ZZ}	$	&	$	1	$	\\	[3pt]
\hline
\hline
\end{tabular}
\end{table*}

\subsubsection{Sensitivities and signals}
\label{Sensitivities}

The above discussion shows that the experiment-independent observables
relevant for studies of the anomaly frequency 
of a trapped electron, positron, proton, or antiproton 
are the 36 quantities
$\btw e J$, $\btws e J$, $\bftw e {(JK)}$, $\bftws e {(JK)}$
and 
$\btw p J$, $\btws p J$, $\bftw p {(JK)}$, $\bftws p {(JK)}$.
In the special case of comparative tests between particles and antiparticles
performed with magnetic fields of the same magnitude $B$,
these observables reduce to the 18 differences 
$\De\btw e J$, $\De\bftw e {(JK)}$
and $\De\btw p J$, $\De\bftw p {(JK)}$.
As a guide to existing and prospective sensitivities
to Lorentz and CPT violation
that could be obtained,
we consider next a subset of sensitive
Penning-trap experiments
measuring the anomaly frequency for these species.

The experiments chosen for the discussion here are listed in 
Table \ref{geometry}.
For each experiment,
we show the species involved,
the colatitude $\ch$ of the laboratory,
the direction $\hat x_3$ of the magnetic field,
and its magnitude $B$.
For completeness and reference,
we also provide the longitude $\la$ of the laboratory
and the offset time $T_0$
relating the local sidereal time $T_\oplus$ to the  
canonical time $T$ in the Sun-centered frame
according to Eq.\ \rf{T0}.

Consider first experiments sensitive to the electron sector.
A 1999 experiment at the University of Washington 
\cite{de99}
compared the anomaly frequencies of electrons and positrons
with a precision of about 2 ppt.
The data were analysed for an effect independent of sidereal time,
so the reported results can be viewed as time-averaged measurements
of the predicted difference \rf{omdiff}.
The cyclotron frequencies used for the two species were almost identical,
so the form \rf{equalomdiff} can be adopted to obtain 
sensitivities to the coefficients $\De\btw e J$, $\De\bftw e {(JK)}$.
Using the geometric factors in Table \ref{geometry}
and the rotation transformation \rf{transf},
the expression \rf{equalomdiff} can be converted to the Sun-centered frame.
The relevant combinations of tilde coefficients
are shown in the first two lines of Table \ref{electronsector}.
We conservatively take the constraint $b \lsim 50$ rad/s 
reported in Ref.\ \cite{de99}
to represent the limit $|b^Z_e| \lsim 5\times 10^{-24}$ GeV
in the present notation.
Averaging the result over time 
and substituting for the values in Ref.\ \cite{de99}
then gives the bound
\bea
&&
\Big| \De\btw e Z 
+ (4\times 10^{-16}{\rm ~GeV}^2) (\De\bftw e {XX} + \De\bftw e {YY})
\hskip 40pt
\nn\\
&&
\hskip 30pt
+ (8\times 10^{-16}{\rm ~GeV}^2) \De\bftw e {ZZ}
\Big |
\lsim 7\times 10^{-24} {\rm ~GeV.}
\nn\\
\label{99-1}
\eea
The result \rf{99-1} can be viewed as generalizing the published result
to the tilde coefficients $\De\btw e Z$ and $\De\bftw e {JJ}$
or, equivalently,
as incorporating also the basic coefficients
$\gcmw 4 {XYT} e$, $\bcw 5 {ZTT} e$, $\gcw 6 {XYTTT} e$, 
$\bcfw 5 {XYZ} e$, $\bcfw 5 {YZX} e$, $\bcfw 5 {ZXY} e$,
$\gcfw 6 {XYTXY} e$,
$\gcfw 6 {YZTYZ} e$,
and $\gcfw 6 {ZXTZX} e$.

\renewcommand{\arraystretch}{1.5}
\begin{table*}
\caption{
\label{protonsector}
Analysis for the proton sector.}
\setlength{\tabcolsep}{5pt}
\begin{tabular}{cccc}
\hline
\hline
Experiment	&		Lab.\ frame		&			Sun-centered frame				&		Harmonic		\\	\hline
ATRAP \cite{di13}	&	$	\btws p 3	$	&	$	-0.7	\btws p X			$	&	$	\codt	$	\\	
	&	$		$	&	$	-0.7	\btws p Y			$	&	$	\sodt	$	\\	
	&	$	\bftws p {33}	$	&	$	-	\bftws p {(XZ)} 			$	&	$	\codt	$	\\	
	&	$	  	$	&	$	-	\bftws p {(YZ)} 			$	&	$	\sodt	$	\\	
	&	$		$	&	$	-0.2	(\bftws p {XX} - \bftws p {YY} )			$	&	$	\ctodt	$	\\	
	&	$		$	&	$	-0.5	\bftws p {XY}  			$	&	$	\stodt	$	\\	[5pt]
BASE \cite{mo14}	&	$	\btw p 3	$	&	$	0.8	\btw p X			$	&	$	\codt	$	\\	
	&	$		$	&	$	0.8	\btw p Y			$	&	$	\sodt	$	\\	
	&	$	\bftw p {33}	$	&	$	-	\bftw p {(XZ)} 			$	&	$	\codt	$	\\	
	&	$	  	$	&	$	-	\bftw p {(YZ)} 			$	&	$	\sodt	$	\\	
	&	$		$	&	$	0.3	(\bftw p {XX} - \bftw p {YY} )			$	&	$	\ctodt	$	\\	
	&	$		$	&	$	0.6	\bftw p {XY}			$	&	$	\stodt	$	\\	[5pt]
BASE \cite{chip}	&	$	\De\btw p 3	$	&	$	0.3	\De\btw p Z			$	&	$	1	$	\\	
	&	$		$	&	$	-0.4	\De\btw p X -	0.9	\De\btw p Y	$	&	$	\codt	$	\\	
	&	$		$	&	$	0.9	\De\btw p X +	0.4	\De\btw p Y	$	&	$	\sodt	$	\\	
	&	$	\De\bftw p {33}	$	&	$	0.4	(\De\bftw p {XX} + \De\bftw p {YY}) +	0.1	\De\bftw p {ZZ}	$	&	$	1	$	\\	
	&	$		$	&	$	-0.2	\De\bftw p {(XZ)} -	0.6	\De\bftw p {(YZ)} 	$	&	$	\codt	$	\\	
	&	$	  	$	&	$	0.6	\De\bftw p {(XZ)} -	0.2	\De\bftw p {(YZ)} 	$	&	$	\sodt	$	\\	
	&	$		$	&	$	-0.3	(\De\bftw p {XX} - \De\bftw p {YY} ) +	0.6	\De\bftw p {(XY)} 	$	&	$	\ctodt	$	\\	
	&	$		$	&	$	-0.3	(\De\bftw p {XX} - \De\bftw p {YY} ) -	0.6	\De\bftw p {(XY)} 	$	&	$	\stodt	$	\\	
\hline
\hline
\end{tabular}
\end{table*}

Another 1999 analysis at the University of Washington
\cite{mi99}
reported results from an analysis of the sidereal variation
of the anomaly frequency of a trapped electron
measured over a two-month period,
with the magnetic field continuously calibrated.
Fitting the data to a sinusoid at the sidereal frequency $\om_\oplus$
and constraining its amplitude
yielded a $2\si$ limit of $|\de\om_a^e| \leq 8\times 10^{-25}$ GeV 
in the present notation.
The combinations of tilde coefficients $\btw e J$ and $\bftw e {(JK)}$
relevant for this experiment
are shown in Table \ref{electronsector}.
These expressions reveal that the experiment 
places the constraint
\bea
&&
\hskip-10pt
\Big(\big[0.7 \btw e X + (10^{-15}{\rm ~GeV}^2)\bftw e {(XZ)}\big]^2 
\nn\\
&&
\hskip 30pt
+ \big[0.7 \btw e Y + (10^{-15}{\rm ~GeV}^2)\bftw e {(YZ)}\big]^2
\Big)^{1/2} 
\nn\\
&&
\hskip 100pt
\lsim 4\times 10^{-25} {\rm ~GeV.}
\label{99-2}
\eea
Comparing this result to the bound \rf{99-1} 
obtained using the 1999 comparison 
of electron and positron anomaly frequencies 
shows that the sidereal analysis 
constrains different spatial components of the tilde coefficients.
Moreover,
CPT-even effects are also contained in the result \rf{99-2}
via $H$- and $d$-type basic coefficients.

A more recent Penning-trap experiment 
at Harvard University 
\cite{ha11}
measured the $g$ factor of the electron
to 0.28 ppt.
This impressive precision offers in principle
improved sensitivity to several components
of the tilde coefficients $\btw e J$, $\bftw e {(JK)}$
via the study of sidereal variations
in analogy to the 1999 work discussed above
\cite{mi99}.
No such analysis has been performed to date,
but we present in Table \ref{electronsector}
the relevant combinations of tilde coefficients
for the first and second harmonics
required for this study
and specific to the geometric factors of the experiment.

A similar sidereal analysis for a trapped positron
could be performed using another experiment under development
at Harvard University 
\cite{fo15}.
This would offer not only the first sensitivity 
to components of the tilde coefficients $\btws e J$, $\bftws e {(JK)}$
but would also permit improved measurements 
of the experiment-independent CPT-odd observables 
$\De\btw e Z$ and $\De\bftw e {JJ}$
by comparison with measurements for the electron.
Table \ref{electronsector} contains
the combinations of tilde coefficients
for the first and second harmonics 
for the planned positron experiment.
We also list the components of the CPT-odd observables
that could be constrained
by comparing the anomaly frequencies for the electron and positron
at the same magnetic field strength.
Note that sidereal variations of these difference coefficients
are also of interest.
The corresponding expressions in the Sun-centered frame 
follow from the definitions \rf{debdefs},
so they appear in the same linear combinations
up to an overall sign for the antiparticle coefficients.

Next,
we consider experiments sensitive to the proton sector.
In an experiment located at CERN,
the ATRAP collaboration has measured
the magnetic moment of the antiproton to 4.4 ppm
\cite{di13}.
In principle,
an analysis of sidereal variations using this experiment 
could yield measurements of some components 
of the tilde coefficients $\btws p J$ and $\bftws p {(JK)}$,
which would represent the first sensitivity achieved
to these physical effects.
The components accessible to the geometry of this experiment
via first and second harmonics of the sidereal frequency
are listed in the first few lines of Table \ref{protonsector}.

A measurement of the magnetic moment of the proton
at the record sensitivity of 3.3 ppb has recently been performed
by the BASE collaboration in an experiment located in Mainz
\cite{mo14}.
In this case,
a search for sidereal variations could in principle 
provide sensitivity to certain components 
of the tilde coefficients $\btw p J$ and $\bftw p {(JK)}$.
Table~\ref{protonsector}
shows the combinations of coefficients
that would be accessible to this experimental geometry.
The BASE collaboration also plans to perform a version 
of this experiment at CERN,
which is at a different colatitude,
using a different orientation and strength of the bore of the primary magnet
and ultimately using a quantum logic readout
that will permit rapid measurements 
of the proton and antiproton anomaly frequencies
\cite{chip}.
This offers the opportunity to measure many components 
of the tilde coefficients 
$\btw p J$, $\bftw p {(JK)}$,
$\btws p J$, and $\bftws p {(JK)}$
via sidereal-variation studies.
The constant parts and the sidereal variations of the differences
$\De\btw p J$ and $\De\bftw p {(JK)}$
would also be measurable with this setup.
Using the geometrical factors listed in Table \ref{geometry}
reveals that this future experiment can access 
the combinations of difference components shown in Table \ref{protonsector}.
Modulo an overall sign for the antiproton case,
the same linear combinations of tilde coefficients appear in sidereal studies, 
as can be deduced by inspection of the definitions \rf{debdefs}.

Taken together,
the published results 
\cite{di13,mo14}
from the ATRAP and BASE experiments can be combined
to extract estimated constraints on experiment-independent observables
for Lorentz and CPT violation.
The methodology to derive these constraints 
is of potential interest for future experiments as well,
so we outline it here.
Consider the comparison \rf{sumdiff},
recalling that all anomaly frequencies
are functions of the form \rf{omalv}.
Since both experiments took data over an extended time period,
we can plausibly approximate the sidereal variations 
as averaging to zero,
leaving only the constant shifts.
This means only the dependence on the colatitude
and on the direction and strength of the magnetic fields
needs to be considered.
For BASE,
the colatitude is $\ch \simeq 40.0^\circ$ 
and the magnetic field $B\simeq 1.9$ T points to local south,
corresponding to the $\hat x$ direction
in the standard laboratory frame 
discussed in Sec.\ \ref{Frame changes}.
For ATRAP,
the colatitude is $\ch^* \simeq 43.8^\circ$ 
and the magnetic field $B^*\simeq 5.2$ T points upwards,
along the $\hat z$ direction in the standard laboratory frame.
The expressions \rf{delomaw} for the frequency shifts
can then be combined to yield
\bea
\De \om_a^{p} &\equiv&
\half (\de \om_a^{p} - \de \om_a^{\ol p})
\nn\\
&=& 
\btw p x - \bftw p {xx} B 
+ \btws p z - \bftws p {zz} B^*
\nn\\
&=& 
- \btw p Z \sin\ch
+ \btws p Z \cos\ch^*
\nn\\
&&
- \half (\bftw p {XX} + \bftw p {YY}) B \cos^2\ch 
- \bftw p {ZZ} B \sin^2\ch 
\nn\\
&&
- \half (\bftws p {XX} + \bftws p {YY}) B^* \sin^2\ch^*
- \bftws p {ZZ} B^* \cos^2\ch^* ,
\nn\\
\Si \om_a^{p} &\equiv&
\half (\de \om_a^{w} + \de \om_a^{\ol w})
\nn\\
&=& 
\btw p x - \bftw p {xx} B 
- \btws p z + \bftws p {zz} B^*
\nn\\
&=& 
- \btw p Z \sin\ch
- \btws p Z \cos\ch^*
\nn\\
&&
- \half (\bftw p {XX} + \bftw p {YY}) B \cos^2\ch 
- \bftw p {ZZ} B \sin^2\ch 
\nn\\
&&
+ \half (\bftws p {XX} + \bftws p {YY}) B^* \sin^2\ch^*
+ \bftws p {ZZ} B^* \cos^2\ch^* .
\nn\\
\eea
These results can be entered on the right-hand side
of the comparison \rf{sumdiff}.
Using the numerical values of the other quantities
reported by the ATRAP and BASE measurements 
and keeping only one significant figure 
in light of the approximations made,
we obtain the 2-sigma limit
\bea
&&\Big| \btw p Z - 0.4 \btws p Z 
\nn\\
&&
\hskip 10pt
+ (2\times 10^{-16}{\rm~ GeV}^{2})(\bftw p {XX} + \bftw p {YY}) 
\nn\\
&&
\hskip 10pt
+ (2\times 10^{-16}{\rm~ GeV}^{2})\bftw p {ZZ} 
\nn\\
&&
\hskip 10pt
+ (1\times 10^{-16}{\rm~ GeV}^{2})(\bftws p {XX} + \bftws p {YY}) 
\nn\\
&&
\hskip 10pt
+ (3\times 10^{-16}{\rm~ GeV}^{2})\bftws p {ZZ} 
\Big|
\lsim 2\times 10^{-21} {\rm ~GeV}.
\qquad
\label{atrapbase}
\eea
This is the desired experiment-independent measure
of Lorentz and CPT violation in the proton sector,
which is specific to the comparison 
of the BASE proton and ATRAP antiproton results. 

\renewcommand{\arraystretch}{1.5}
\begin{table}
\caption{
\label{epconstraints}
Constraints on tilde coefficients.}
\setlength{\tabcolsep}{5pt}
\begin{tabular}{ccc}
\hline
\hline
		Coefficient			&			Constraint			&	Ref.	\\	\hline
$	|	\btw e X	|	$	&	$	<	6\times 10^{-25}	{\rm ~GeV}	$	&	\cite{mi99}	\\	
$	|	\btw e Y	|	$	&	$	<	6\times 10^{-25}	{\rm ~GeV}	$	&	\cite{mi99}	\\	
$	|	\btw e Z	|	$	&	$	<	7\times 10^{-24}	{\rm ~GeV}	$	&	\cite{de99}	\\	
$	|	\btws e Z	|	$	&	$	<	7\times 10^{-24}	{\rm ~GeV}	$	&	\cite{de99}	\\	
$	|	\bftw e {XX} + \bftw e {YY}	|	$	&	$	<	2\times 10^{-8}	{\rm ~GeV}^{-1}	$	&	\cite{de99}	\\	
$	|	\bftw e {ZZ}	|	$	&	$	<	8\times 10^{-9}	{\rm ~GeV}^{-1}	$	&	\cite{de99}	\\	
$	|	\bftw e {(XZ)} 	|	$	&	$	<	4\times 10^{-10}	{\rm ~GeV}^{-1}	$	&	\cite{mi99}	\\	
$	|	\bftw e {(YZ)} 	|	$	&	$	<	4\times 10^{-10}	{\rm ~GeV}^{-1}	$	&	\cite{mi99}	\\	
$	|	\bftws e {XX} + \bftws e {YY}	|	$	&	$	<	2\times 10^{-8}	{\rm ~GeV}^{-1}	$	&	\cite{de99}	\\	
$	|	\bftws e {ZZ}	|	$	&	$	<	8\times 10^{-9}	{\rm ~GeV}^{-1}	$	&	\cite{de99}	\\	[5pt]
$	|	\btw p Z	|	$	&	$	<	2\times 10^{-21}	{\rm ~GeV}	$	&	\cite{di13,mo14}	\\	
$	|	\btws p Z	|	$	&	$	<	6\times 10^{-21}	{\rm ~GeV}	$	&	\cite{di13,mo14}	\\	
$	|	\bftw p {XX} + \bftw p {YY}	|	$	&	$	<	1\times 10^{-5}	{\rm ~GeV}^{-1}	$	&	\cite{di13,mo14}	\\	
$	|	\bftw p {ZZ}	|	$	&	$	<	1\times 10^{-5}	{\rm ~GeV}^{-1}	$	&	\cite{di13,mo14}	\\	
$	|	\bftws p {XX} + \bftws p {YY}	|	$	&	$	<	2\times 10^{-5}	{\rm ~GeV}^{-1}	$	&	\cite{di13,mo14}	\\	
$	|	\bftws p {ZZ}	|	$	&	$	<	8\times 10^{-6}	{\rm ~GeV}^{-1}	$	&	\cite{di13,mo14}	\\	
\hline
\hline
\end{tabular}
\end{table}

Each of the three constraints
\rf{99-1},
\rf{99-2},
and \rf{atrapbase}
obtained above 
involves several tilde coefficients.
Some intuition for the scope of these constraints
can be obtained by assuming each coefficient in turn
to be the only nonzero one and determining its bound.
This procedure, 
which is common practice across many subfields
searching for Lorentz and CPT violation
\cite{tables},
neglects any cancellation or interference
among different coefficients
but does offer insight and a reasonable measure 
of the sensitivity to individual coefficients
provided no signal has been observed.
The resulting constraints on each tilde coefficient
are displayed in Table \ref{epconstraints}.
All 16 of these bounds are new in detail
because they include effects from $d=4$, 5, and 6
that are analyzed for the first time in the present work.
As described above,
some of them reduce in an appropriate limit
to results already reported in a suitable minimal-SME limit.
Note that a large number of the 36 independent observables
remain unexplored in Penning-trap experiments to date.

\section{Summary and outlook}
\label{Summary}

In this work,
we explore the prospects for searching for Lorentz- and CPT-violating
effects using experiments with Penning traps.
We begin in Sec.\ \ref{Theory}
by presenting the Lagrange density 
for Lorentz-violating spinor QED 
with operators of mass dimensions up to six.
The minimal-SME terms in this theory
are given in Eqs.\ \rf{lag3} and \rf{lag4},
while the complete set of terms at $d=5$ and 6 
is displayed in
Eqs.\ \rf{l5d}, \rf{l5f}, \rf{l6d}, \rf{l6f}, and \rf{l6df}.
The basic properties of the corresponding coefficients
for Lorentz violation
are compiled in Table \ref{nrcoeff}.

Determining the observables in the theory
requires investigating the interplay between different operators  
under field redefinitions.
We perform a general fermion field redefinition \rf{redef}
and list the resulting transformations
in Table \ref{fieldredefs}.
Among other results,
this analysis demonstrates that many terms in the Lagrange density
that couple spinors to the electromagnetic field strength
can be absorbed into other terms in the theory
via suitable field redefinitions.
A result of practical utility in this work 
involves the case of a constant electromagnetic field,
for which the piece \rf{l6df} of the Lagrange density vanishes,
while all the $F$-type coupling terms in Eqs.\ \rf{l5f} and \rf{l6f}
can be generated by the replacements \rf{replacement} and \rf{repl2}
in the Lagrange-density terms \rf{l5d} and \rf{l6d}.

Another issue in characterizing the observables
for Lorentz and CPT violation
is the noninertial nature of any laboratory 
on the surface of the Earth.
In Sec.\ \ref{Frame changes},
we discuss three relevant frames for experimental analysis:
the inertial Sun-centered frame,
the standard noninertial laboratory frame,
and a noninertial apparatus frame. 
Allowing for the rotation of the Earth,
the transformations required to achieve the inertial Sun-centered frame
are given by Eqs.\ \rf{rotmat} and \rf{euler}.
This analysis neglects the suppressed boost effects 
arising from the revolution of the Earth about the Sun,
which would be an interesting subject for a separate work.
 
Applications of the theory to experiments with Penning traps
are discussed in Sec.\ \ref{Application to Penning traps}.
We use perturbation theory to determine the effects
of Lorentz and CPT violation 
on the relativistic Landau levels of a particle 
in a uniform magnetic field.
The results obtained are at leading order in Lorentz violation
but exact in other quantities.
They are found to be lengthy and are presented in the Appendix.
The dominant Lorentz- and CPT-violating perturbative shifts
of the energy levels are given in Eq.\ \rf{deleb},
while the corresponding results for antiparticles
are presented in Eq.\ \rf{delebbar}.
These expressions permit the derivation of the
dominant Lorentz- and CPT-violating shifts 
of the cyclotron and anomaly frequencies
of trapped particles and antiparticles.
At leading order,
the cyclotron-frequency shifts \rf{cyclnochange}
are found to vanish.
The leading-order shifts in anomaly frequencies
for particles and antiparticles of species $w$
are given explicitly by Eq.\ \rf{delomaw}.
We use the latter expressions to show that
the difference \rf{Deltaom} between these anomaly frequencies 
is a measure of pure CPT violation
in idealized comparative experiments
with the same orientation and magnitude
of the trapping magnetic field,
while the sum \rf{Sumom} 
involves only CPT-even effects.

Turning next to issues closer to experiment,
we discuss observable signals 
for trapped electrons, positrons, protons, and antiprotons.
Since the anomaly frequency \rf{omalv} depends
on the magnitude of the magnetic field 
and on geometric factors including the local sidereal time,
the colatitude of the experiment,
and the local direction of the magnetic field,
it follows that the ratio of the anomaly to cyclotron frequencies
is no longer an intrinsic property of the particle
but becomes an experiment-dependent quantity.
We prove that
the intrinsic observables providing experiment-independent measures 
of Lorentz and CPT violation are instead the 36 tilde coefficients
$\btw e J$, $\btws e J$, $\bftw e {(JK)}$, $\bftws e {(JK)}$
and 
$\btw p J$, $\btws p J$, $\bftw p {(JK)}$, $\bftws p {(JK)}$.
Comparisons of results for particles and antiparticles
must also take this into account.
One consequence is that 
the difference \rf{sumdiff} between
the ratios of the anomaly to cyclotron frequencies
for a particle and an antiparticle
typically contains both CPT-odd and CPT-even effects.

The above results make feasible the analysis of existing and future
experiments for sensitivity to experiment-independent observables
for Lorentz and CPT violation.
The theory predicts oscillations of all observables
at specific harmonics of the sidereal frequency,
along with time-independent signals 
that can be detected in comparative experiments
with particles and antiparticles.
To illustrate the methodology for the analysis,
we consider the sensitive experiments listed in Table \ref{geometry}
and examine some of their implications.
The key information permitting the extraction of constraints
on observables is derived and tabulated
for electrons and positrons in Table \ref{electronsector} 
and for protons and antiprotons in Table \ref{protonsector}.
Existing experimental measurements 
are used to extract new and improved constraints 
on numerous tilde coefficients for Lorentz and CPT violation,
using the sidereal variation of observables 
and comparisons between particles and antiparticles.
In the electron sector,
we obtain the bounds \rf{99-1} and \rf{99-2}
using results from experiments at the University of Washington
\cite{de99,mi99},
while in the proton sector we combine
independent results from the ATRAP \cite{di13}
and BASE \cite{mo14} experiments
to obtain the bound \rf{atrapbase}.
Table \ref{epconstraints} lists the ensuing 16 constraints
obtained when a single tilde coefficient
is taken to be nonzero at a time.

We close this work with a brief outlook on some open and feasible projects
that would further enhance the role of Penning traps
in studying the foundational Lorentz and CPT symmetry of nature.
Each of the following five general topics represents
an open challenge for theory and experiment,
the resolution of which will ultimately require 
disentangling conceptual and calculational issues
and performing analyses to extract constraints from experimental data.

{\it 1.\ Boost effects.}
A comparatively direct extension of the present work
would involve investigation of suppressed effects neglected here.
The largest of these effects comes from the revolution
of the Earth about the Sun,
which introduces harmonics of the annual revolution frequency
and corresponding sidebands near the sidereal harmonics.
The new observables come with a suppression factor
of the Earth's boost $\be_\oplus\simeq 10^{-4}$,
but they include coefficient combinations
that are unobservable without the boost.
Additional smaller effects associated with the boost of the laboratory
due to the rotation of the Earth,
which are suppressed by $\be_L\simeq 10^{-6}$,
are also of potential interest.
While boosts can generate sensitivity to coefficients
otherwise unobservable in Penning-trap experiments,
the corresponding shifts in the Landau levels
remain independent of the level number,
so much of the conceptual structure for the treatment of signals 
given in the present work remains in force.
The techniques for handling the boosts
have been developed in several prior contexts
\cite{gkv14,kv15,ca04,he08,space}
and could be transferred to Penning-trap analyses. 

{\it 2.\ Cyclotron-frequency shifts.} 
Qualitatively different suppressed effects
arise from subleading Lorentz- and CPT-violating contributions 
to the energy shifts that are proportional to $|qB|$.
These contributions can be extracted 
from the expressions for the energy shifts
for $d\leq 5$ given in Appendix \ref{Energy shift}.
The suppression factors are stronger than those for boosts,
being of order $10^{-9}$ for electrons or positrons
and of order $10^{-16}$ for protons or antiprotons.
However,
many of the contributions produce energy shifts 
that depend on the level number,
so they can change the relative spacing
of the lowest-lying levels in a single Landau stack 
and hence affect the cyclotron frequency
as well as the anomaly frequency.
This implies that signals for Lorentz and CPT violation
can appear not only in measurements 
of anomalous magnetic moments
but also in measurements of charge-to-mass ratios. 
Signals of this type have been studied theoretically in the minimal SME 
\cite{bkr98},
and they have led to constraints using
experiments comparing the cyclotron frequencies 
of antiprotons and H$^-$ ions
\cite{ga99,ul15}.
Revisiting the theoretical basis for these works 
while including effects at $d=5$ and 6 
can be expected to yield interesting new constraints
and stimulate further experiments.

{\it 3.\ Field effects.} 
Additional sensitivities to Lorentz and CPT violation
could be obtained by refining the analysis of the electromagnetic fields
in a realistic Penning trap.
For example,
the presence of the electric field
that restricts the axial motion of the trapped particle
produces several types of novel and potentially interesting effects.
If the experimental procedure equipartitions
the axial and cyclotron energies,
then effects from the axial motion will be comparable 
to those proportional to $|qB|$ mentioned above
and so will permit suppressed sensitivities
to additional coefficients for Lorentz and CPT violation
associated with covariant-derivative couplings
of the trapped fermion.
These kinds of effects correspond to terms involving powers of $p_3$ 
in the energy shifts given in Appendix \ref{Energy shift}.
The electric field also introduces new sensitivities
to the $F$-type coefficients,
which are associated with electromagnetic couplings of the fermion
that vanish in the absence of the electric field.  
For a constant electric field,
these effects can be derived from
the energy shifts given in Appendix \ref{Energy shift}
by performing the replacements \rf{replacement} and \rf{repl2}.
Moreover,
the spatially varying electric field in a realistic Penning trap
could offer sensitivity to the terms 
in the Lagrange density \rf{l6df}
that otherwise are inaccessible.
Control of the magnetic field also implies interesting prospects
for studying independent observables.
For example,
the dependence on $B$ of the anomaly frequency \rf{omalv}
shows that two experiments differing only 
in the magnitudes of the magnetic fields 
can yield sensitivities to coefficients for Lorentz violation. 

{\it 4.\ Other species.} 
Trapping and studying the magnetic moments of species 
other than electrons, positrons, protons, and antiprotons
could provide additional sensitivities 
to coefficients for Lorentz and CPT violation
beyond those discussed here.
For example,
experiments on any ion with magnetic moment 
influenced by the neutrons in its nucleus could offer sensitivities 
to coefficients in the neutron sector.
A theoretical treatment of this possibility
along the lines in the present work
would make an interesting project
with the potential to influence experimental discovery.
The coefficients for Lorentz violation
for composite species are combinations
of those in the electron, proton, and neutron sectors,
and determining the relationship 
a crucial part of this type of investigation.
For H$^-$ ions in the context of the minimal SME,
the link has been established at leading order in Lorentz violation
and shown to imply experimental sensitivities
differing from those for trapped electrons or protons
\cite{bkr98}.
Inclusion of operators with $d=5$, 6 would introduce
unique dependences on the momenta of the particles
forming the composite species.
For nuclear components with comparatively high momenta,
this implies a potential increase in the experimental reach
by several orders of magnitude,
in line with results from atomic spectroscopy
\cite{kv15}.

{\it 5.\ Additional SME sectors.} 
Efforts to extend the theoretical scope of our analysis 
can also be expected to provide interesting and novel results.
One option would be to extend the results in this work
to operators of arbitrary $d$.
Partial results along these lines are given in Eq.\ \rf{allorders}.
Another line of investigation 
would consider effects from other SME sectors. 
For example,
contributions from Lorentz and CPT violation in the photon sector
are known to modify the Maxwell equations
and hence could in principle affect the behavior of trapped particles,
although most effects are tightly constrained by other 
experiments
\cite{tables,photonexpt,photonastro}.
Effects on trapped particles from Lorentz and CPT violation 
in the strong, electroweak, or gravitational sectors
could also be envisaged.
Many of these are likely to be suppressed in typical scenarios.
For example,
effects proportional to 
the local gravitational acceleration in the laboratory
must come with a numerical suppression factor of order $10^{-32}$.
In light of the current reach of experiments with Penning traps,
countershaded Lorentz and CPT violation
\cite{kt09}
may be the most interesting possibility to pursue in this context.

In conclusion,
this work presents the general theory for Lorentz- and CPT-violating QED
including operators of mass dimensions $d\leq 6$
and offers a guide to the prospects
for detecting dominant effects from Lorentz and CPT violation
in precision experiments on particles and antiparticles
confined to a Penning trap.
We have used the methodology developed here and existing experimental data 
to constrain 16 of the 36 experiment-independent observables
for Lorentz and CPT violation in the electron and proton sectors,
but much work remains before a complete coverage
of all predicted dominant effects can be achieved.
The many prospective effects 
in current and future Penning-trap experiments
provide strong inducement for continuing
these types of efforts to investigate Lorentz and CPT symmetry,
with promising potential for uncovering violations
of these basic spacetime symmetries.

\section*{Acknowledgments}

This work was supported in part
by the Department of Energy under grant number {DE}-SC0010120
and by the Indiana University Center for Spacetime Symmetries.

\onecolumngrid

\appendix
\section{Perturbative energy shifts}
\label{Energy shift}

In this Appendix,
we present the results of perturbative calculations
for the energy levels of a fermion
of mass $m$, charge $q=\si |q|$, and spin orientation $s=\pm 1$.
The analysis is performed 
using Eq.\ \rf{deltae} with Lorentz- and CPT-violating operators 
appearing in
$\cl^{(3)}$, $\cl^{(4)}$, and $\cl^{(5)}_D$.
As discussed in Sec.\ \ref{Perturbative energy shift},
the contributions from $\cl^{(5)}_F$ 
can be obtained 
via the substitutions presented in Sec.\ \ref{Scenario}.
The expressions below are valid at leading order in Lorentz violation
but are otherwise exact. 
They are presented in the apparatus frame 
having coordinates $(x^1, x^2, x^3)$
described in Sec.\ \ref{Frame changes},
with the magnetic field aligned along $\hat x_3$.

At $d=3$,
calculation with $\cl^{(3)}$
reveals contributions to the energy shift given by
\bea
\de \ens^{(3)} &=&
a^0
- a^3 \fr {p_3} {\ens}
+ \Z s b^0 \fr{p_3}{\ens} 
- \Z s b^3\left(1 - \fr{(2n+1 - \Z s )|qB|}{\ens(\ens + m)}\right)
- \Z s H^{12}\left(1 - \fr{p_3^2}{\ens(\ens + m)}\right).
\eea
For $d=4$,
we obtain from $\cl^{(4)}$
the results
\bea
\de \ens^{4} &=&
-c^{00}\ens
+(c^{03}+c^{30})p_3
-(c^{11}+c^{22})\fr{(2n+1-\si s)|qB|}{2\ens}
-c^{33}\fr{p_3^2}{\ens}
\nn\\
&&
-\si s d^{00}p_3
+\si s d^{30}m\left(1-\fr{p_3^2}{\ens(\ens+m)}\right)
+\si s(d^{03}+d^{30})\fr{p_3^2}{\ens}
\nn\\
&&
-\si s(d^{11}+d^{22})p_3\fr{(2n+1-\si s)|qB|}{2\ens(\ens+m)}
-\si s d^{33}p_3\left(1-\fr{(2n+1-\si s)|qB|}{\ens(\ens+m)}\right)
-e^{0}m+e^{3}p_3\fr{m}{\ens}
\nn\\
&&
+\si s g^{120}\left(m+\fr{(2n+1-\si s)|qB|}{\ens+m}\right)
-\si s g^{123}p_3\left(\fr{m}{\ens}+\fr{(2n+1-\si s)|qB|}{\ens(\ens+m)}\right)
\nn\\
&&
+\si s (g^{231}-g^{132})p_3\fr{(2n+1-\si s)|qB|}{2\ens(\ens+m)}
+\si s (g^{012}-g^{021})\fr{(2n+1-\si s)|qB|}{2\ens}.
\eea
The contributions from $\cl^{(5)}_D$ at $d=5$ are found to be
\bea
\de \ens^{(5)} &=&
m^{(5)00}m\ens
-2m^{(5)03}p_3m
+(m^{(5)11}+m^{(5)22})|qB|
\left(
\fr {(2n+1)m}{2\ens}+\si s \fr {(2n+1-\si s)|qB|}{2\ens(\ens+m)}
\right)
+m^{(5)33}p_3^2\fr{m}{\ens}
\nn\\
&&
+a^{(5)000}{\ens}^2
-2a^{(5)003}p_3\ens
+(a^{(5)011}+a^{(5)022}) 
   \left(\fr{2n+1}{2}-\si s \fr{(2n+1-\si s)|qB|}{2\ens(\ens+m)}\right)|qB|
\nn\\
&&
+a^{(5)033}p_3^2
+(a^{(5)101}+a^{(5)202})(2n+1-\si s)|qB|
-(a^{(5)113}+a^{(5)223})p_3\fr{(2n+1-\si s)|qB|}{\ens}
\nn\\
&&
-a^{(5)300}p_3\ens+2a^{(5)303}p_3^2
-(a^{(5)311}+a^{(5)322})p_3\fr{(2n+1)|qB|}{2\ens}
-a^{(5)333}p_3^3\fr{1}{\ens}
\nn\\
&&
+\si s b^{(5)000}p_3\ens
-2 \si s b^{(5)003}p_3^2
+\si s (b^{(5)011}+b^{(5)022})p_3\fr{(2n+1)|qB|}{2\ens}
+\si s b^{(5)033}p_3^2\fr{1}{\ens}
\nn\\
&&
+ \si s (b^{(5)101}+b^{(5)202})p_3\fr{(2n+1-\si s)|qB|}{(\ens+m)}
- \si s (b^{(5)113}+b^{(5)223})p_3^2\fr{(2n+1-\si s)|qB|}{\ens(\ens+m)}
\nn\\
&&
-\si s b^{(5)300}\ens\left(\ens-\fr{(2n+1-\si s)|qB|}{\ens+m}\right)
+2 \si s b^{(5)303}p_3\left(\ens-\fr{(2n+1-\si s)|qB|}{\ens+m}\right)
\nn\\
&&
-\si s (b^{(5)311}+b^{(5)322})|qB|
\left(
\fr{2n+1}{2}+\si s \fr{(2n+1-\si s)|qB|}{\ens(\ens+m)}
\right)
-\si s b^{(5)333}p_3^2\left(1-\fr{(2n+1-\si s)|qB|}{2\ens(\ens+m)}\right)
\nn\\
&&
+\si s (H^{(5)0102}-H^{(5)0201})(2n+1-\si s)|qB|
-\si s (H^{(5)0123}-H^{(5)0212})p_3\fr{(2n+1-\si s)|qB|}{\ens}
\nn\\
&&
-\si s H^{(5)1200}{\ens}^2\left(1-\fr{p_3^2}{\ens(\ens+m)}\right)
+2 \si s H^{(5)1203}p_3\left(\ens-\fr{p_3^2}{\ens+m}\right)
\nn\\
&&
-\si s (H^{(5)1211}+H^{(5)1222})|qB|
\left(
\fr {(2n+1)m}{2\ens}+\fr {(2n+1-\si s)^2|qB|}{2\ens(\ens+m)}
\right)
-\si s H^{(5)1233}p_3^2\left(1-\fr{p_3^2}{\ens(\ens+m)}\right)
\nn\\
&&
+\si s (H^{(5)1302}-H^{(5)2301})p_3\fr{(2n+1-\si s)|qB|}{(\ens+m)}
-\si s (H^{(5)1323}-H^{(5)2313})p_3^2\fr{(2n+1-\si s)|qB|}{\ens(\ens+m)}.
\eea
The corresponding energy shifts for the antiparticle
can be obtained as described 
in Sec.\ \ref{Perturbative energy shift}.

\twocolumngrid

\end{document}